\documentclass[preprint,superscriptaddress,showpacs,aps,pre,floatfix,longbibliography]{revtex4-1}
\usepackage[utf8]{inputenc} 
\usepackage[T1]{fontenc}    
\usepackage{booktabs}       
\usepackage{amsfonts}       
\usepackage{nicefrac}       
\usepackage{microtype}      
\usepackage{amssymb,amsthm}
\usepackage{amsmath,bm}
\usepackage{graphicx,epstopdf}
\usepackage{multirow}
\usepackage{subfigure}
\usepackage{setspace}
\usepackage{wrapfig}
\usepackage{comment}
\usepackage{listings}
\usepackage{amssymb,dsfont}
\usepackage{leftidx}
\usepackage{bigdelim}
\usepackage{mathrsfs}
\usepackage{blkarray,adjustbox}
\usepackage{lineno}
\usepackage{enumitem}
\usepackage{mathpazo}
\usepackage{rotating}
\usepackage{soul}
\usepackage[dvipsnames]{xcolor}
\usepackage{xr-hyper}
\usepackage[pdftex, pdftitle={Article},colorlinks, pdfauthor={Author}]{hyperref} 
\usepackage[nameinlink,capitalise]{cleveref}
\definecolor{darkgray}{gray}{0.35}
\hypersetup{colorlinks={true},linkcolor={darkgray},citecolor=ForestGreen}

\usepackage{xr}
\makeatletter
\newcommand*{\addFileDependency}[1]{
  \typeout{(#1)}
  \@addtofilelist{#1}
  \IfFileExists{#1}{}{\typeout{No file #1.}}
}
\makeatother

\newcommand*{\myexternaldocument}[1]{
    \externaldocument[si:]{#1}
    \addFileDependency{#1.tex}
    \addFileDependency{#1.aux}
}
\myexternaldocument{SI}

\crefname{section}{Sec.}{Secs.}
\Crefname{section}{Section}{Sections}
\crefname{figure}{Fig.}{Figs.}
\Crefname{figure}{Figure}{Figures}
\crefname{equation}{Eq.}{Eqs.}
\Crefname{equation}{Equation}{Equations}

\makeatletter

\renewcommand*{\p@subsection}{}

\renewcommand*{\p@subsubsection}{}
\makeatother

\definecolor{reddish}{HTML}{FBB4AE}
\definecolor{blueish}{HTML}{B3CDE3}
\definecolor{magentish}{HTML}{FF00AA}
\definecolor{greenish}{HTML}{a1d99b}

\newcommand{\todo}[1]{}
\renewcommand{\todo}[1]{{\color{orange}[[TODO: {#1}]]}}

\graphicspath{{./figs/}} 

\begin{document}
\title{Contrasting social and non-social sources of predictability in human mobility}

\author{Zexun Chen}
    \affiliation{BioComplex Laboratory, University of Exeter, UK}
    \affiliation{College of Engineering, Mathematics and Physical Sciences, University of Exeter, UK}
\author{Sean Kelty}
    \affiliation{Department of Physics and Astronomy, University of Rochester, Rochester, NY, USA}
\author{Brooke Foucault Welles}
    \affiliation{Northeastern University, Boston, MA, USA}
\author{James~P.~Bagrow}
	\affiliation{Department of Mathematics \& Statistics and Vermont Complex Systems Center, University of Vermont, Burlington, VT, USA}
\author{Ronaldo Menezes}
    \affiliation{BioComplex Laboratory, University of Exeter, UK}
    \affiliation{College of Engineering, Mathematics and Physical Sciences, University of Exeter, UK}
\author{Gourab Ghoshal}
    \email[Correspondence: ]{gghoshal@pas.rochester.edu}
    \affiliation{Department of Physics and Astronomy, University of Rochester, Rochester, NY, USA}
    \affiliation{Department of Computer Science, University of Rochester, Rochester, NY, USA}


\begin{abstract}
\begin{singlespace}
Social structures influence a variety of human behaviors including mobility patterns, but the extent to which one individual's movements can predict another's remains an open question.
Further, latent information about an individual's mobility can be present in the mobility patterns of both social and non-social ties, a distinction that has not yet been addressed.
Here we develop a ``colocation'' network to distinguish the mobility patterns of an ego's social ties from those of non-social colocators, individuals not socially connected to the ego but who nevertheless arrive at a location at the same time as the ego. 
We apply entropy and predictability measures to analyse and bound the predictive information of an individual's mobility pattern and the flow of that information from their top social ties and from their non-social colocators.
While social ties generically provide more information than non-social colocators, we find that significant information is present in the aggregation of non-social colocators: 
3--7 colocators can provide as much predictive information as the top social tie, and colocators can replace up to 85\% of the predictive information about an ego, 
compared with social ties that can replace up to 94\% of the ego's predictability.
The presence of predictive information among non-social colocators raises privacy concerns: given the increasing availability of real-time mobility traces from smartphones, individuals sharing data may be providing actionable information not just about their own movements but the movements of others whose data are absent, both known and unknown individuals.
\end{singlespace}
\end{abstract}


\maketitle

\section{Introduction}

The recent availability of extensive geolocated datasets related to human movement, has enabled the quantitative study of human movement at an unprecedented level~\cite{Barbosa2018}, contributing greatly to insights in estimating migratory flows, traffic forecasting, urban planning, mitigating pollution and epidemic modeling among other applications~\cite{Batty2013,simini_2012,Uherek2010,Lee2017,Kirkley_2018,Pan_2013,tizzoni_2012_real,toole_2015_path}. 
Several common regularities have been observed across these studies, including bursty activity rates, tendencies to visit a select few locations disproportionately more than others, as well as decreasing likelihood to explore as time goes on~\cite{Vazquez2006,Brockmann2006, Song2010a,Rhee2011a, Boyer2012, Hasan2013a}. A related aspect that can enhance the potential of these findings, particularly for urban planning and the control of epidemics, is the ability to predict the future locations of individuals or groups using their prior history of travel. 
Indeed, it has been shown that a perfect algorithm can predict, with between 70-90\% certainty, an individual's future location given their prior location visits~\cite{Song_2010}, depending upon the spatiotemporal granularity of observations~\cite{Ikanovic_2017}.

Human beings are also typically highly social creatures and social structures can influence behavior in a variety of human activities including movement patterns.
In fact, it has been shown that social relationships statistically account for between 10\% to 30\% of all human movement~\cite{Cho2011}.
Social structures inherently encode information flow between parties, such that residual information about an individual can be inferred from their social ties~\cite{Aral:2017jo}.
Such a phenomenon was demonstrated in the context of online interactions, where about 95\% of an individual's potential predictive accuracy was contained in their social network, despite no recourse to information about the person in question~\cite{bagrow2019information}.
Coupled with the observation that movement patterns in the virtual and physical domains are strikingly similar ~\cite{Hazarie2020}, this leads to the intriguing question as to whether one can leverage a person's social network to predict their future mobility patterns, absent any information on their own history.
This possibility holds promise for a number of applications, and may be particularly relevant in the context of mitigating pandemics ~\cite{Zhou_2020,Wu_2020}, where a key tool in the arsenal is contact tracing based on  mobility patterns ~\cite{Flaxman_2020, Davies_2020, Oliver:2020lq}. However, accurately mapping human mobility can be challenging due to understandable privacy concerns and people's willingness to disclose or share personal data~\cite{Fahey_2020, Bengio_2020}.

Location-Based Social Networks (LBSNs) yield opportunities to examine social relations to human mobility, containing information both about sequences of location visits and (in some cases) information on the underlying social network. 
At the same time, spatially aggregating these data can reveal
individuals in different social circles who visit similar or overlapping locations; for instance, people working in the same building but with different companies, or parents whose children attend the same schools but are unknown to each other.
These non-social ties are potential predictors of a person's mobility trajectory. 
Terming such individuals ``non-social colocators,'' we ask whether and to what extent such colocators yield predictive mobility information, and how this information compares to that of social ties.

Here we apply non-parametric information-theoretic estimators to study human mobility extracted from three Location-Based Social Networks (LBSNs), that contain sequences of location trajectories as well as the (reported) social network of a subset of users. 
We demonstrate the existence of information transfer in these networks, finding that a given ego's future location visits can be predicted, with between 80--100\% of the ego's own accuracy, by studying the historical patterns of just 10 of their alters (ranked by number of common locations visited).  
Remarkably, non-social colocators, while individually providing less information than social ties, can in the aggregate provide similar levels of predictability. The information flow provided by colocators is also surprisingly robust to temporal-displaced colocations, implying users that never physically colocate can still provide comparable information to social ties. Indeed, the information transfer appears to be driven by the overlap of unique locations visited by the ego and alters, in both social and non-social colocators.

The rest of this paper is organized as follows.
\Cref{sec:pre-processing} describes the mobility and social datasets we study.
We apply information-theoretic tools to these data in \cref{sec:results}, where we quantify the predictive information of individuals (\cref{sec: individual_information}), their social ties and their non-social colocators (\cref{sec: cross_information}).
We examine the spatial and temporal underpinnings of our results (\cref{sec: spatial_temporal}) and end in \cref{sec:discussion} with a discussion of the implications of our findings.

\section{Materials}\label{sec:pre-processing}

Our study uses three publicly available datasets that contain both mobility traces and the social network of a subset of the users of the platforms. The first is BrightKite, a location-based social networking service (LBSN)~\cite{Cho2011, Grabowicz2013} containing  4,491,143 geo-tagged check-ins by 58,228 users over a period of Apr 2008 - Oct 2010. 
The second dataset is from Weeplaces, a website that  generated visualizations and reports from location-based check-ins in platforms such as Facebook and Foursquare~\cite{Hazarie2020}. The considered data contains only Foursquare check-ins, that includes 7,658,368 geotagged check-ins produced by 15,799 users from Nov 2003 to Jun 2011. 
Finally, we also consider Gowalla~\cite{Cho2011}, another LBSN consisting of 6,442,890 check-ins by 196,591 users over a period of Feb. 2009 - Oct. 2010. 
(For more details, see Sec. S1.) 

Within these data, 
each \textit{event}, i.e., an instance of a location visit, is timestamped and tagged with a unique location ID.
In all datasets, a location visit $v$ is represented by a tuple $v = (u,\ell,t)$, meaning a user $u$ visited a location $\ell$ at time $t$.
At a user level, a trajectory composed of $N_u$ discrete observations is characterized by a sequence of $N_u$ location-time pairs $(\ell_{i},t_{i})_{i \in 1\ldots N_u}$ where $\ell_{i}$ stands for the location visited at step $i$ and time $t_{i}$.
We assume a user who visits $N_u$ locations in total, visits $n_u \leq N_u$ \textit{distinct} locations, with equality holding only if the user never visits a location more than once.
To filter out spurious activities, we exclude inactive users and discard records with missing attributes. Furthermore, for purposes of statistical significance, we also discard users who have logged $N_u < 150$ check-ins (our results are robust to these filtering criteria; see Sec. S1.3 and Figs. S3,S4).
After filtering, we are left with 
510,308 events across 6,132 users in Brightkite, 
924,666 events across 11,533 users in Weeplaces, and finally 
850,094 events across 9,937 users for Gowalla 
(cf. Tab. S1 for further details).  
In Fig. S1 we show the check-in maps for each of the datasets indicating global coverage with the highest concentrations in North America and Western Europe. 
In Fig. S2 we plot the corresponding total distinct locations visited by all users, jump-length and radius of gyration distributions finding statistical trends in the LBSN data consistent with other sources of mobility data~\cite{Barbosa2018}.

Each of the datasets have social networks collected by their respective API's (details in Sec. S1.1), however, not all of the users in the network log check-ins. Given that our goal is to examine the information transfer in these social networks as it relates to location visits, we focus on users with logged location-trajectories. 
To quantify the information provided by colocated non-social ties, we construct colocation networks where a tie is included between an ego and alter if they checked in at the same location within a particular time window (See Sec. S2 for details on egocentric network construction). We assume that individuals who colocate more often contain more predictive information about one other's whereabouts, so the ranking criteria is based on the frequencies of colocations (Sec. S2.4, Figs. S5, S6).
All results presented in the main manuscript correspond to a one hour temporal bin, but our results are robust to varying temporal frames (Sec. S2.5 and Fig. S7).

\section{Results}\label{sec:results}

\subsection{Information contained in egos}\label{sec: individual_information}

\begin{figure}[t!]
 	\centering
 	\includegraphics[width=0.8\linewidth]{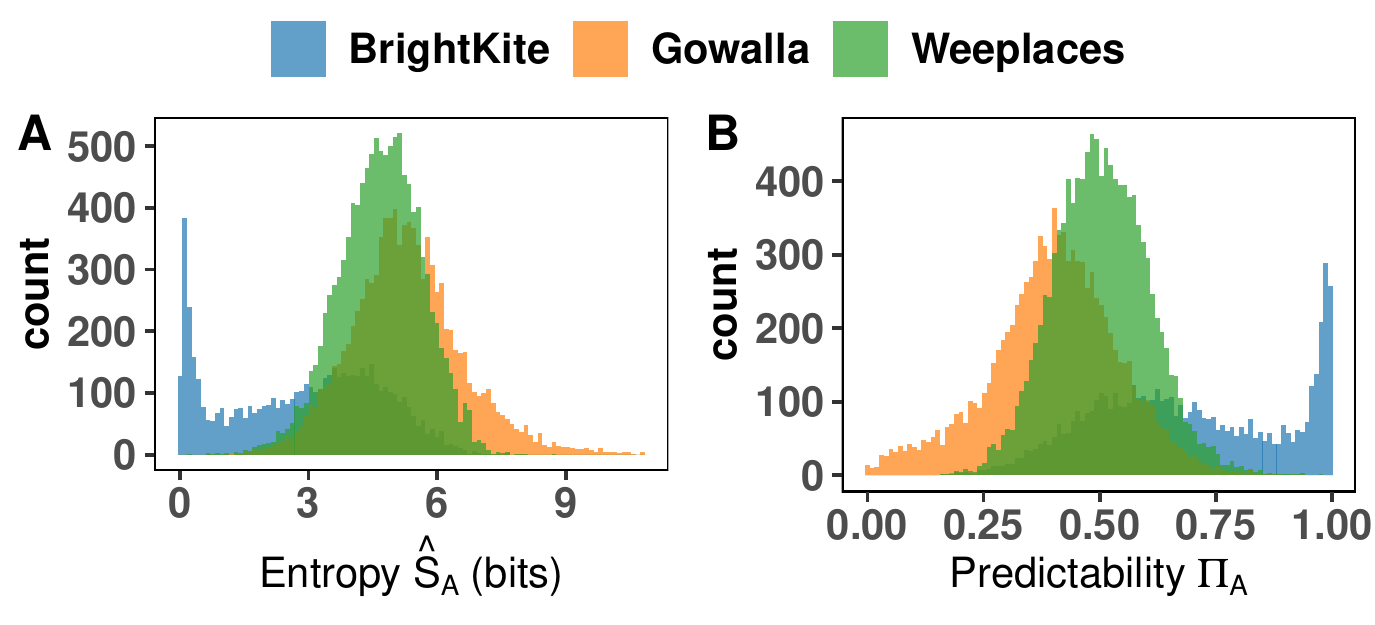}
 	\caption{ \textbf{Entropy and Predictability in three mobility datasets.} 
 	\textbf{A},
 	The distribution of the entropy $\hat{S}_A$ (\cref{eq:lempel}) for each of the three datasets. 
 	\textbf{B},
 	The corresponding distribution of predictability $\Pi_A$, calculated by inverting \cref{eq:predict}, tells us how well an ideal algorithm can predict an individual's future location given their mobility history.
 	}
 	\label{fig:dataset_S_Pi}
\end{figure}

We begin our analysis by examining the information contained in the location trajectories of all ego's in each of the datasets; this serves as a baseline when comparing information flow with social and non-social alters. 
The degree of uncertainty in capturing the future locations of a trajectory $A$, given past observations, is encoded in the entropy rate $S_A$ of the trajectory. 
Accounting for both frequency of location visits, as well as temporal ordering (specific ordered sequences in the data), we make use of a non-parametric estimator~\cite{Lempel1976,Kontoyiannis1998} given by the expression
\begin{equation}
    \hat{S}_A= \frac{N \log_2 N}{\sum\limits_{i=1}^N\Lambda_{i}},
    \label{eq:lempel}
\end{equation}
where for a trajectory $A$ of $N$ moves of an individual,
$\Lambda_{i}$ is the length of the shortest trajectory sub-sequence beginning at position $i$ not seen previously, and the entropy is measured in bits. 
This estimator has been applied to mobility patterns and online social activities~\cite{Song_2010,bagrow2019information}.
In the absence of any structure in the sequence, the expression converges to the standard Shannon entropy~\cite{bagrow2019information}. 
In \cref{fig:dataset_S_Pi}{\bf A}, we plot $\hat{S}_A$ for the three datasets finding peaks between 4--5 bits with varying degrees of spread. (One dataset, BrightKite, looks distinct from the others for reasons we discuss shortly.)

An intuitive measure to interpret these results is the \textit{perplexity} $2^{S}$:
we are as uncertain about future visits for a trajectory with entropy rate 3 bits, for example, as we would be when choosing uniformly at random from $2^{3}=8$ possible locations.
Using $\hat{S}_A$ from Fig.~\ref{fig:dataset_S_Pi}{\bf A}, this implies that on average knowing the past history of the typical ego allows us to reduce the possible number of future location visits to between 16--32 sites. 
Given the average number of distinct locations (Fig. S2) that a typical user visits across the datasets (107 total distinct locations per user on average in BrightKite, 166 in Weeplaces, and 198 in Gowalla), information due to the spatiotemporal regularities of ego trajectories represent an order of magnitude reduction from choosing across all locations uniformly at random, a result consistent with that found in other mobility studies~\cite{Song_2010, Ikanovic_2017}.

The entropy rate can also be interpreted using Fano's inequality~\cite{Cover_2006} to define the \textit{predictability} $\Pi_A$, the upper bound of how often an \textit{ideal predictive algorithm} can correctly guess the next location visit, given prior history. 
This predictability is calculated by inverting the expression,
\begin{equation}
\hat S_A \leq H_A(\Pi_A) + (1 - \Pi_A)\log_{2}(n-1),
\label{eq:predict}
\end{equation}
where $n$ is the number of distinct locations visited and $H(x)$ is the binary entropy function 
capturing the entropy of a simple Bernoulli trial (in this case achieving maximal predictability or not). 
Utilizing $\Pi_A$ allows us to leverage information theory to mathematically bound the performance of all real predictive methods given an information sources inferred uncertainty.

\Cref{fig:dataset_S_Pi}{\bf B} shows the distributions of predictability, finding key difference across the three platforms. 
One dataset in particular, BrightKite, shows a distinct spike of highly predictable ($\Pi_A\approx 1$) users.
In BrightKite, a large fraction of users visit only between 1--3 unique locations (Fig. S2{\bf A}), leading to a low entropy rate and thus a disproportionately high degree of predictability.  
Conversely, in Gowalla, while $\Pi_A$ is peaked at $\approx 40$\%, we find a wide spread around the peak, given that some users visit many locations (indeed, 23 users never return to a previously visited location), leading to a high entropy rate, and low predictability. 
This likely stems from Gowalla incentivizing its users to discover new locations (see Sec. S1). 
Finally, in Weeplaces, $\Pi_A$ is peaked at $\approx50$\% with a tighter bound around the peak as compared to the other datasets. 
This observed diversity in mobility behavior across the three platforms provides us with a robustness check on the results to follow.

\subsection{Information contained in social ties and non-social colocators}
\label{sec: cross_information}

Next we examine information flow, how much mobility information about the ego is contained in the sequence of location visits of their alter(s), absent any information about the ego's own location history. 
This can be measured by the \textit{cross-entropy}~\cite{bagrow2019information, Ziv_1993}, which is greater than the entropy when the alter contains less information on the ego than the ego itself, and quantifies information loss when we have access to only the alter's past. 
To estimate the cross-entropy between two sequences $A$ and $B$ (representing the mobilities of the ego and alter, respectively), Eq.~\eqref{eq:lempel} can be modified according to
\begin{equation}
    \hat{S}_{A|B} = \frac{N_A \log_2(N_B)}{\sum_{i=1}^{N_A}\Lambda_i(A|B)},
 \label{eq:CE}
  \end{equation}
where $N_A$ and $N_B$ are the lengths of the sequences $A,B$, and the cross-parsed match length $\Lambda_i(A|B)$, is the length of the shortest location sub-sequence starting at position $i$ of sequence $A$ not previously seen in sequence $B$. 
Here, `previously' refers to those locations $\ell_j$ in sequence $B$ with $t_j< t_i$, the timestamp of the check-in location $\ell_i$ in  sequence $A$. 
As with the cross-entropy, 
one can generalize the predictability $\Pi_A$ to the cross-predictability $\Pi_{A|B}$ by applying \cref{eq:predict} to \cref{eq:CE}.
For the remainder of this paper, both social and colocated non-social alters have been processed by retaining alters that provide better-than-random information about their ego, as well as removing from the colocation network any spurious colocators (see Sec. S2 and Tab. S2).

\begin{figure}[t!]
	\centering
	\includegraphics[width=0.99\linewidth]{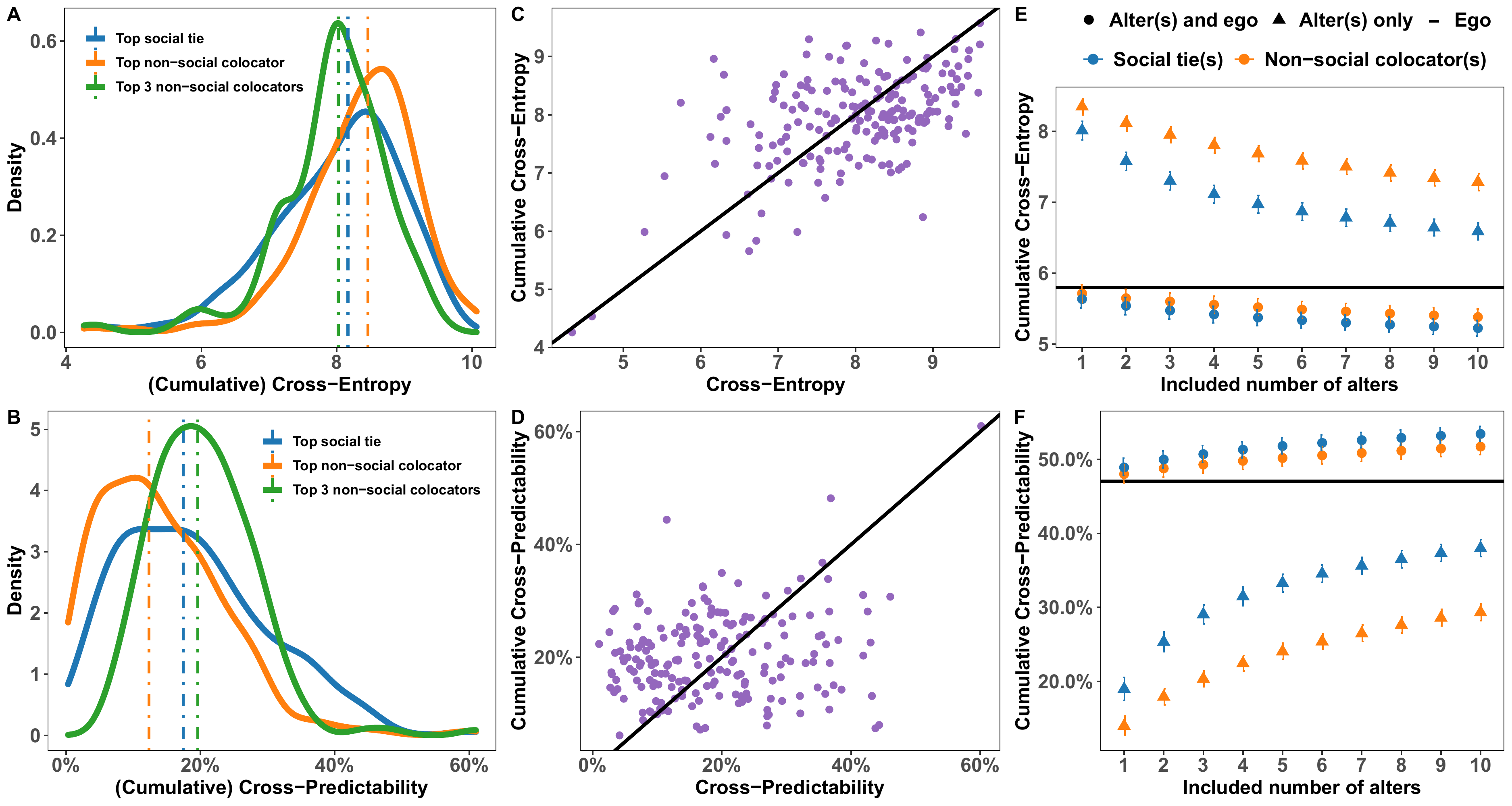}
	\caption{\textbf{Cross-entropy and predictability in social ties and non-social colocators}.
	\textbf{A} 
	Distributions of $\hat S_{A|B}$ for the rank-1 social tie (median 8.17 bits), non-social colocator (median 8.46 bits), and $\hat S_{A|\mathcal{B}}$ for the top-3 non-social colocators (median 8.02 bits) in Weeplaces.  
	\textbf{B}
	The corresponding $\Pi_{A|B}$ for the social  (median17.43\%), and non-social colocators (median 12.35\%), and  $\Pi_{A|\mathcal{B}}$ for the top-3 non-social colocators (median 19.60\%).  
	\textbf{C}
	$\hat S_{A|B}$ encoded in the top-social tie as a function of $\hat S_{A|\mathcal{B}}$ for the top-3 non-social colocators. Each point corresponds to a single ego and the solid line denotes $y = x$.
	\textbf{D}
	As in panel {\bf C} but with predictability instead of cross-entropy.
	\textbf{E, F} 
	$\hat S_{A|\mathcal{B}}$ and $\Pi_{A|\mathcal{B}}$ after accumulating the top-ten social alters and non-social colocators. 
	Horizontal lines denote the average entropy ($5.80$ bits) of egos and their self-predictability ($47.05$\%). Shapes indicate whether the self-predicatbility of the the ego was included in the sequence (circles) or excluded (triangles).
	Error bars denote $95\%$ CI.
	}
	\label{fig:wp_CCP_social_vs_non_social}
\end{figure}

\Cref{fig:wp_CCP_social_vs_non_social} shows the results of our information metrics on the Weeplaces dataset. 
Panels {\bf A} and {\bf B} show the distribution of the cross-entropy and predictability for the rank-1 social tie and non-social colocated alter. 
We see that the top social tie provides slightly more information than the top colocator, with predictability slightly right-skewed (\cref{fig:wp_CCP_social_vs_non_social}{\bf B}). 
While social ties provide more predictive information, the distribution also shows the existence of some non-social colocators that provide mobility information comparable to that provided by social ties. Furthermore, the predictability of egos are positively correlated with the predictability of their top alter (Fig. S10), meaning highly predictable egos tend to have highly predictable top alters, and similarly more unpredictable egos tend to have less predictable alters. These findings are consistent with that seen for Brightkite (Fig. S8{\bf A,B}) and Gowalla (Fig. S9{\bf A,B}).

We've thus far looked at the individual and pair-wise information in ego-alter pairs, a limited analysis, given that these are being considered as information sources in isolation. Next, we examine the information content of a multiplicity of an ego's alters, or in other words examine the information content of a subset of the ego's colocators by adapting the cross-entropy to a set of alters.
To estimate the amount of information needed to encode the next location of sequence $A$ given the location information in \textit{a set of sequences} $\mathcal{B}$, we generalize the pairwise cross-entropy to the \textit{cumulative cross-entropy} according to \cite{bagrow2019information}, thus, 
\begin{equation}\label{eq:CCE}
    \hat{S}_{A| \mathcal{B}} = \frac{N_A \log_2(N_{A\mathcal{B}})}{ \sum_{i=1}^{N_A}\Lambda_i(A|\mathcal{B})},
\end{equation}
where $\Lambda_i(A|\mathcal{B})  = \max \{\Lambda_i(A|B), B \in \mathcal{B} \}$ is the longest cross-parsed match length over any of the sequences in the set of sequences $\mathcal{B}$, $N_{A\mathcal{B}} = \sum_{B \in {\mathcal{B}}}w_B N_B / \sum_{B \in {\mathcal{B}}}w_B$ is the average of the lengths $N_B$, and $w_b$ is the number of times that matches from sequence $A$ are found in each sequence $B \in \mathcal{B}$. 
Note that if there is only one sequence in $\mathcal{B}$, Eq.~\eqref{eq:CCE} reduces to 
Eq.~\eqref{eq:CE}. 
By applying Fano's inequality (Eq.~\eqref{eq:predict}), we denote the corresponding cumulative cross-predictability as $\Pi_{A| \mathcal{B}}$.
In Fig. S6 we see that on average, alters associated with more frequent colocations contain more information content, as a consequence, we rank alters according to frequency of colocations. Plotting the average number of colocations between ego-alter pairs saturate at around 10 alters in all three datasets (Fig. S5), and therefore 
we examine the information content of the top-10 most frequently colocated social alters and  non-social colocated alters.
For a fair comparison between these two different sources of mobility information, we focus on egos in both the social and colocation network with at least ten alters in each network, leading to 33 (Brightkite), 97 (Gowalla), and 199 (Weeplaces) egos (cf.~Tab. S2).

By moving from one non-social colocator to three, we see in \cref{fig:wp_CCP_social_vs_non_social}{\bf A} and {\bf B} that considerably more predictive information is present, with the peak of $\Pi_{A|B}$ shifted significantly rightward. Further, the peak is now at a higher value than the peak for the top social tie (\cref{fig:wp_CCP_social_vs_non_social}{\bf B}), indicating that many egos are better predicted by three non-social colocators than they are by their top social tie.
We further emphasize this relationship in \cref{fig:wp_CCP_social_vs_non_social}{\bf C,D} with scatter plots comparing the cross-entropy of the top social tie to the cumulative cross-entropy of the top-3 non-social colocators; any points above the line $y=x$ in panel {\bf D} demonstrate more information flow from the colocators about the ego than from the top social tie.
Individually, non-social colocators are less informative than social ties, but collectively they can meet or exceed the information content of individual social ties.


Expanding on the comparison between social ties and non-social colocators,
in \cref{fig:wp_CCP_social_vs_non_social}{\bf E,F}, we plot the cumulative cross-entropy and cross-predictability $\Pi_{A| \mathcal{B}}$, finding a progressive increase in predictability as we accumulate more alters (positive Spearman's $\rho$ across 88.94\% all users, $p < 0.05$). 
A given number of social ties provides more information on average than the same number of non-social colocators, as demonstrated by the lower curve in entropy in panel {\bf E} and higher curve in predictability in panel {\bf F}.
Specifically, 94.47\% of egos in Weeplaces show significantly higher social tie predictability than non-social colocator predictability (paired one-sided $t$-test, $p  < 0.01$). 
However, while the colocator curve in \cref{fig:wp_CCP_social_vs_non_social}{\bf F} sits below the social curve for a given number of alters, we do see that on average a greater number of colocators can exceed the information content of a small number of social ties. 
For instance, the top-3 non-social colocators provide higher predictability than the top social tie, and the top-7 colocators provide higher predictability than the top-2 social ties. 
Consistent results hold for the BrightKite and Gowalla datasets (Figs. S8,S9)
and for different definitions of colocation (Fig. S7).

While alter information about the ego is important to understand, especially for matters of privacy (see Discussion), we also wish to understand whether that information is redundant when given the ego's past. 
We therefore show in \cref{fig:wp_CCP_social_vs_non_social} {\bf E,F} curves including the ego's past alongside that of the alters. 
We find that non-redundant information exists in both types of alters, with a gain of $\approx 10\%$ predictability for the top-10 social ties and $\approx 14\%$ for the top-10 non-social colocators.
\Cref{fig:wp_CCP_social_vs_non_social}{\bf F} also demonstrates that $\Pi_{A|\mathcal{B}}$ appears to saturate as more alters are included, an effect observed in other studies~\cite{bagrow2019information}.
This saturation effect is examined in Sec. S4, where $\Pi_{A|\mathcal{B}}$ is extrapolated beyond our data window of 10 alters by fitting a nonlinear saturating function and estimating the extrapolation predictability $\Pi_\infty$.
For Weeplaces we find  $\Pi_{\infty} = 44.32\%$ and $39.79\%$ for social ties and colocators, respectively.
Compared to the average $\Pi_\mathrm{ego} = 47.05\%$ all egos in the network, this means that $94\%$ and $85\%$, respectively, of the \textit{potential} predictability of an ego is in principle available in that ego's alters. The closeness of these values underscore the high degree of predictive information available in the non-social colocators. The corresponding findings for the other two datasets are shown in Tab. S3.

\begin{figure}[t!]
	\centering
	\includegraphics[width=0.7\linewidth]{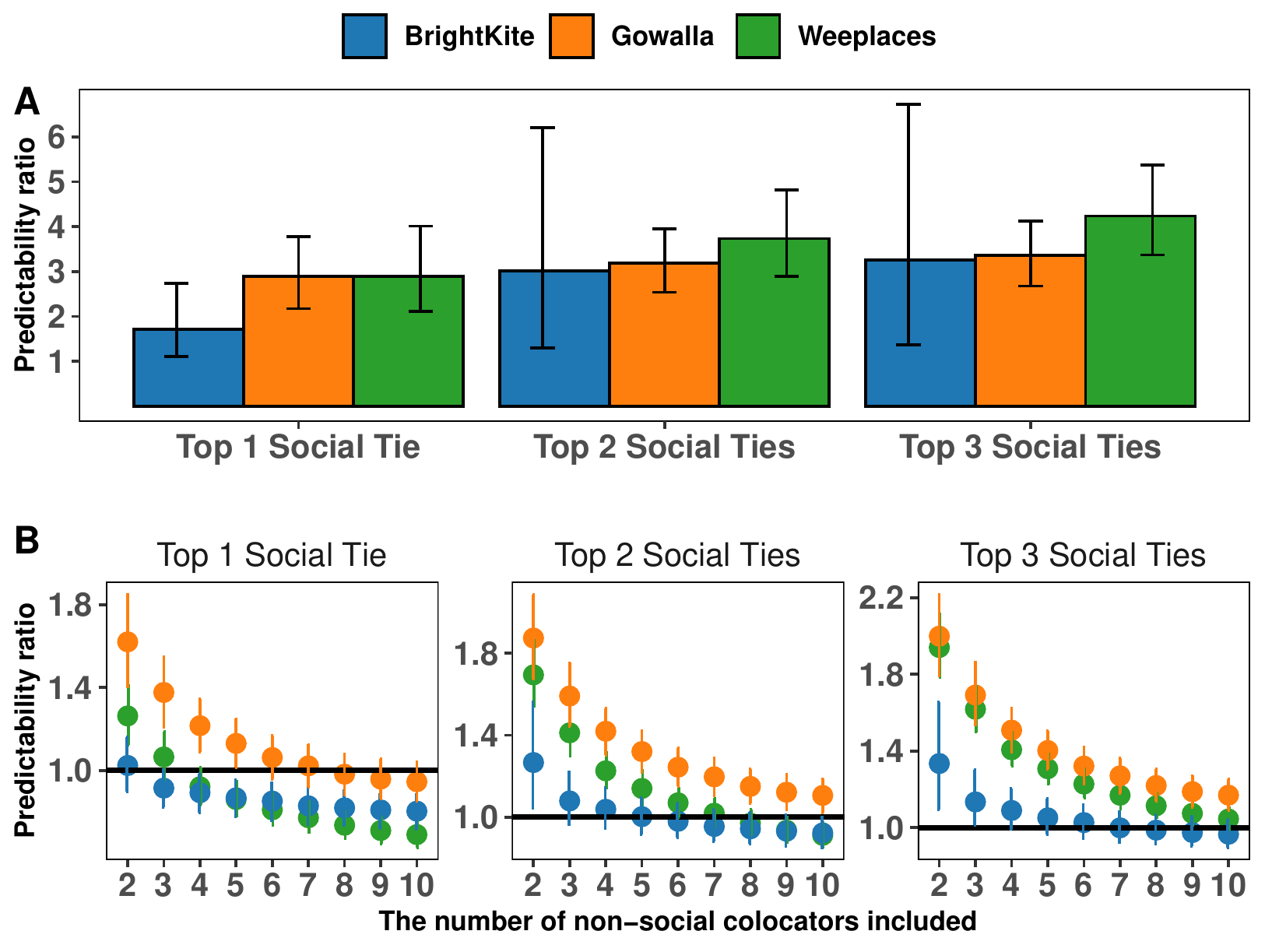}
	\caption{ \textbf{Quantifying the predictive information aggregated from non-social colocator(s) with respect to social ties.}
	\textbf{A}
	The predictability ratio $\Pi_{\text{ego|social tie(s)}}/ \Pi_{\text{ego|non-social colocator(s)}}$ between the top non-social colocator and (left-to-right) the top, top-2, and top-3 social tie(s).
	\textbf{B}
	The predictability ratios between the top 2--10 non-social colocators and the top, top-2, and top-3 social tie(s).
	Error bars denote $95\%$ CI.}
	\label{fig:predictability_ratio}
\end{figure}

Extrapolation analysis demonstrates the overall relative value of non-social colocators, but it does not allow us to determine more precisely how many non-social colocators equal the information content for a given number of social ties.
Therefore, to better quantify the relative information content provided by social ties compared with non-social colocators, we  examine the predictability ratio $\Pi_{\text{ego|social tie(s)}}/ \Pi_{\text{ego|non-social colocator(s)}}$ across all three mobility datasets.
In \cref{fig:predictability_ratio}{\bf A} we present the distributions of predictability ratio comparing the top non-social colocator to the top-$k$ social ties ($k=1,2,3$). 
For the top social ties ($k=1$) we see that BrightKite colocators provide the closest information with a ratio just below 2, meaning the social tie provides approximately twice the predictability of the colocator. In Gowalla and Weeplaces, the difference is even stronger, with the top social ties providing approximately three times the predictability of the top colocator.
Moving from the top colocator to multiple colocators,
in \cref{fig:predictability_ratio}{\bf B} we plot the predictability ratio for increasing numbers of nonsocial colocators; when this curve crosses the horizontal line at a ratio of 1, we have equal amounts of information.
For example, examining the first panel in {\bf B}, we see that curve for BrightKite intersects between 1 and 2 colocators, between 7 and 8 colocators, and between 3 and 4 for Weeplaces. This suggests that three Weeplaces colocators are equivalent to the top social tie, while 7--8 Gowalla colocators are equivalent to the top social tie. 
In Brightkite, a dataset characterized by high predictability and low entropy, one added social tie provides the same degree of information as two non-social colocators.
Across all datasets, we see that an aggregate of fewer than 10 non-social colocators can equal the information of the top social tie.
While more colocators are needed to equal the aggregate of the top-2 or top-3 social ties, the observed decreasing trends suggest a convergence in the amount of information contained in either flavor of tie.

\subsection{Underlying Spatial and Temporal Mobility Characteristics}
\label{sec: spatial_temporal}




We next determine the key factors that determine the near identical types of information transfer in both types of ties, despite having no overlap in the pair-wise connections. One of the possibilities driving the quantity of information on the ego provided by alters is the information inherent in the locations themselves.
That is, it is reasonable to surmise that information about the ego is derived from shared visits to common locations, given that predictability of the ego itself depends on the patterns of location visits in their trajectory.
While alters do not necessarily visit all the locations that their egos do, nor would they necessarily visit at the same time (see below), one can hypothesize that higher-ranked alters share more distinct locations with the ego than lower-ranked ones.
If the trend is similar across both social ties and non-social colocators, then this would be a plausible mechanism for the similarity in the observed cross-predictability.

\begin{figure}[t!]
	\centering
	\includegraphics[width=0.88\linewidth]{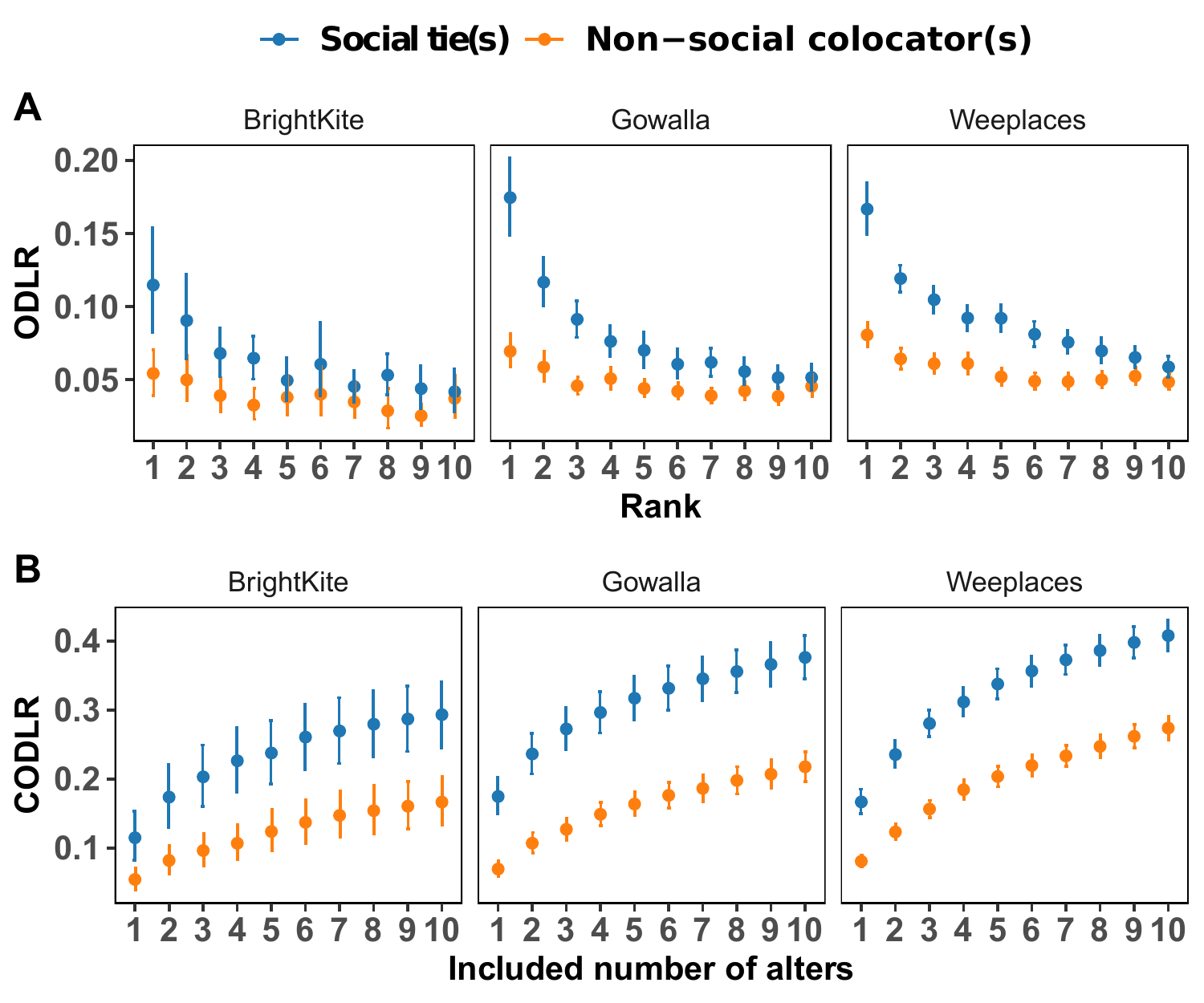}
	\caption{ \textbf{The degree of ego--alter distinct location overlap}.  
	\textbf{A},
	The Overlapped Distinct Location Ratio (\cref{eq:eta1}) indicates that higher ranked alters share more unique location visits than lower-ranked ones, with the top (rank-1) alter showing the most shared location.
	The trend is stronger for social ties than non-social colocators.
	\textbf{B}, 
	The Cumulative Overlapped Distinct Location Ratio (\cref{eq:eta2}) shows increasing discovery of unique locations in the the ego's trajectory as alters are added in order of decreasing rank, but that the rate of discovery slows. 
	Error bars denote $95\%$ CI.
	}
	\label{fig:vip_CCP_VS_ODLR_CODLR}
\end{figure}

To measure this, we compute the proportion of unique locations visited by the ego and its alters. 
For an ego $A$ we define the \textit{Overlapped Distinct Location Ratio (ODLR)} $\eta$ as the fraction of $A$'s visited locations also visited by an alter $B$. Formally,   
\begin{equation}
    \eta_{A|B} = \frac{\left|Y_A \cap Y_B\right|}{\left|Y_A\right|}
    \label{eq:eta1}
\end{equation}
where $Y_A$ and $Y_B$ are the sets of locations visited by $A$ and $B$, respectively, and $\left|\cdot\right|$ denotes set cardinality. 

In \cref{fig:vip_CCP_VS_ODLR_CODLR}{\bf A} we plot the ODLR as a function of alter rank. 
As alters are ranked according to the frequency of overlap of any location visit of the ego, as opposed to distinct location visits, there is no reason to \emph{a priori} expect that a rank-1 alter will share the most number of distinct locations in their trajectory with the ego.
Nevertheless, that is indeed what is observed across all datasets, with a monotonically decreasing trend of ODLR as one considers lower-ranked alters.
This monotonic trend is considerably stronger for social ties than for non-social colocators, although the difference diminishes with the number of alters added.   

\begin{figure}[t!]
	\centering
	\includegraphics[width=0.9\linewidth]{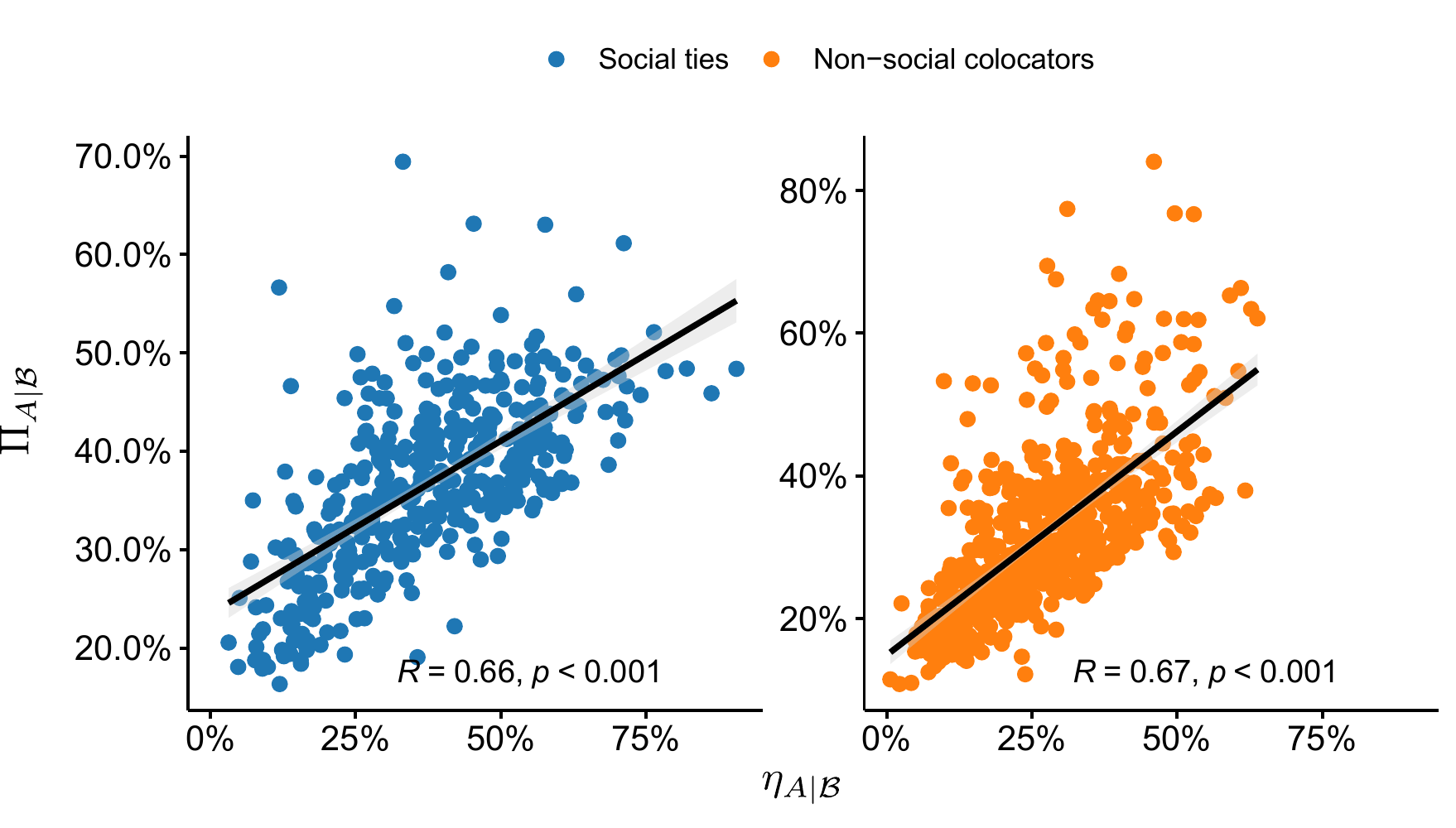}
	\caption{ \textbf{Connecting location overlap to information transfer.}
	Regression analysis of cumulative cross-predictability
	$\Pi_{A|\mathcal{B}}$ and CODLR $\eta_{A|\mathcal{B}}$ for the Weeplaces dataset with the top-10 alters. 
	Here $R$ is Pearson's correlation coefficient. 
	The solid lines are linear regressions.
	}
	\label{fig:vip_MeetupNp_correlation_wp}
\end{figure}

The ODLR fails to consider locations shared across multiple alters, instead focusing on one alter at a time.
Yet we previously saw (\cref{fig:wp_CCP_social_vs_non_social}) the importance of examining the aggregate information, particularly when comparing non-social colocators to social ties.
Therefore, we generalize the ODLR to a \emph{Cumulative Overlapped Distinct Location Ratio} (CODLR) by taking the union of the location sets of multiple alters according to,
\begin{equation}
    \eta_{A|\mathcal{B}} = \frac{|\cup_{B \in \mathcal{B}} (Y_A \cap Y_B)|}{|Y_A|},
    \label{eq:eta2}
\end{equation}
where $A$ is the ego and $\mathcal{B}$ is the set of all alters. 
We plot the results in \cref{fig:vip_CCP_VS_ODLR_CODLR}{\bf B} finding similar increasing monotonic trends across datasets and networks. 
As alters are added, more information on the ego's unique locations are discovered, saturating at between 30-40\% after 10 social ties, and between 15-30\% for non-social colocators. 
Once again we observe that larger numbers of colocators provide comparable location overlap as a smaller number of social ties, emphasizing both the relative importance of social ties and the extent of useful information present in the aggregation of non-social colocators.

We connect ODLR and information flow directly in \cref{fig:vip_MeetupNp_correlation_wp}, by plotting $\eta_{A|\mathcal{B}}$ against $\Pi_{A|\mathcal{B}}$ for the top-10 alters in both types of networks, observing a strong, approximately linear trend (Pearson's $ R\approx 0.66$ for social ties; $R\approx 0.67$ for non-social colocators; both significant, $p<0.001$).
Disentangling the plots by progressively adding alters from rank-1 to rank-10 shows a monotonically increasing trend for the correlations (Figs. S11 and S12). The corresponding results for BrightKite (Figs. S13 and S14) and Gowalla (Figs. S15 and S16) reveal the same trends.

\begin{figure}[t!]
	\centering
	\includegraphics[width=\linewidth]{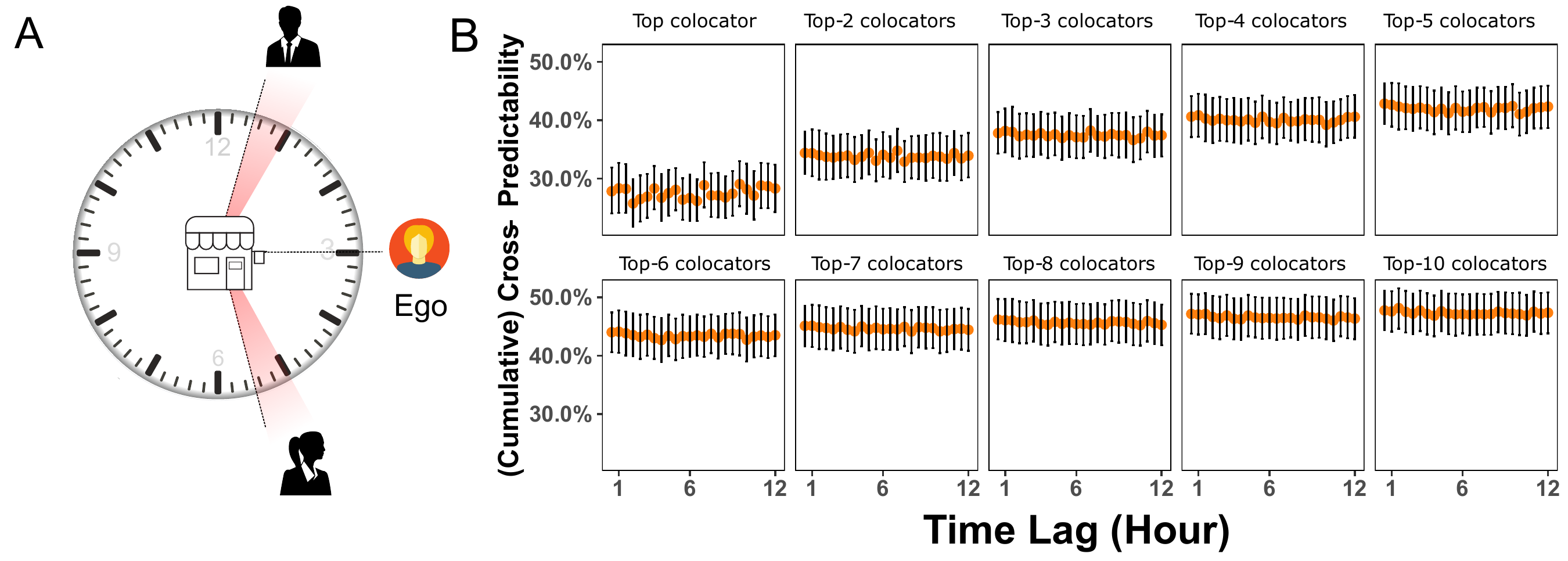}
	\caption{ \textbf{Temporal stability of non-social colocator information}. 
	\textbf{A}
	An example of two time-displaced colocators who visit the same location as the ego on a $T = 2.5$ hour time lag.
	\textbf{B}
	The (cumulative) cross-predictability influence of temporal-lag for non-social colocator(s) in Weeplaces. Each point corresponds to a co-location network resulting from the amount of temporal offset between an ego and alters visit to a common location.
	Error bars denote $95\%$ CI.
	}
	\label{fig:time_lag_effect}
\end{figure}

The observed connection between information transfer and location overlap, behooves one to to ask whether temporal effects are a key factor. In other words, our choice of colocation is based on the simple idea that individuals in the same place at the same time contain information about mobility patterns of each other. 
We can relax this condition and also consider individuals that visit the same locations as the ego, but displaced in time. 
For example, residents of a neighborhood can stop at their local corner store at different times of day and never run into each other, but their visits are always five hours apart because of their respective work schedules. 
We can investigate this by creating networks of time-displaced colocators, where now an ego and alter colocate if an alter visits a location in the time windows $[T,T-1/2]$ hours prior or [$T-1/2,T]$ hours following an ego visitation at the same location (see the illustration in \cref{fig:time_lag_effect}{\bf A}). Note that $T=1/2$ h yields the fully connected window $[-1/2,1/2]$.

We examine the trend in $\Pi_{A|\mathcal{B}}$ as a function of the temporal-lag $T$ in \cref{fig:vip_MeetupNp_correlation_wp}{\bf B}. Each network resulting from the different temporal lags will generally have different ego-alter pairs, and the set of egos with at least 10 alters may change. We consider then the common alters who have at least 10 alters for all the networks between 1/2 h and 12 h with 30-minute intervals. The ego-alter overlap between these networks shown in Fig. S17 indicates that the networks are quite different, meaning co-located alters are not necessarily the same as time-displaced co-located alters.
As before, there is an increase in predictability as alters are added, yet for a given set of alters, remarkably there is little-to-no difference between any of the networks in terms of their information content, at least within the investigated temporal ranges. Furthermore, this suggests potentially unexpected sources of mobility information, as the non-social co-locators selected here do not necessarily have to be visitors of the same location at the same time to provide predictive mobility information about the ego.

\section{Discussion}\label{sec:discussion}

Using information-theoretic measures, we analyzed the spatiotemporal structure of the mobility trajectories of a set of users in three publicly available location-based social networks. 
Entropy measures were used to quantify the sequential information contained in a user's physical trajectory which revealed
differences in our datasets based on the context of how users used the apps. 
Using these measures, we then compared the information present in the mobility patterns of an individual's (the ego's) social ties compared with non-social colocators, other users who frequently visited the same locations as the ego.
Across datasets we found the importance of social ties: consistently more information about the ego's future location was present in the past locations of the social ties than in the past locations of the non-social ties, and this held when aggregating information from multiple number of alters.
Interestingly however, this implies something important: that groups of many non-social colocators can in principle provide as much information as a smaller set of social ties, meaning that non-social sources of mobility information are in principle available.
If access to social data are limited, these non-social data may, in the aggregate, be used as a replacement.


Our study relied on observational data taken from three location-based social networks. 
This introduces crucial caveats. 
In particular, the social ties reported in the datasets are incomplete reflections  of a person's full social circle, and the nature of such ties may differ in the online and offline domains. 
Likewise, not all locations visited by an individual are  recorded in these social networks, which rely primarily on user check-ins, so we expect mobility trajectories to be under-sampled as well.
Followup work, including richer, more detailed data and even experimental studies, are needed to address these concerns, yet our robustness checks, including observing consistent trends across datasets and across our sampling criteria, already provide evidence that our results appear to be robust.
%


The presence of predictive information, both socially and otherwise, has crucial implications.
Privacy protections regarding social data are important to protect sensitive information about a user and their social ties~\cite{pellungrini2017data,Pellungrini:2018vn}. 
Social information flow suggests that an individual's future movements can be predicted by studying the mobility patterns of a few acquaintances. 
On the other hand, our study also demonstrates that social ties are not the only source of predictive mobility information, and measures of colocation are enough to uncover novel sources of mobility information. 
This means that locations monitoring individual visits, for example, a grocery store tracking the smartphones of shoppers~\cite{enck2014taintdroid}, may in principle be collecting the building blocks of mobility profiles, and individuals providing access to their mobility data may also be providing information about both social and non-social ties~\cite{horvat2012,sarigol2014online,garcia2017leaking}. 
While these data can inform important applications such as contact tracing in the early stages of a disease outbreak, significant ethical concerns surrounding such information sources make it critical to place strong access constraints on mobility information~\cite{Waniek:2019mz}. Indeed, the results presented here provide further impetus to the ongoing debate on best practices for privacy protection, both in terms of  legislation and ethical algorithmic development.
\raggedbottom

\bibliographystyle{naturemag}
\bibliography{reference.bib}

\end{document}


%
%

\makeatletter
\renewcommand{\maketitle}{\bgroup\setlength{\parindent}{0pt}
	\begin{flushleft}
{\huge Supporting Information}\\[0.3cm]
		{\large\textbf \@title}
		
		\@author
		
	\end{flushleft}\egroup
}
\makeatother
\title{Contrasting social and non-social sources of predictability in human mobility}

\newcommand{\floor}[1]{\lfloor #1 \rfloor}

\author{Zexun Chen, Sean Kelty, Brooke Foucault Welles, James P.~Bagrow, Ronaldo Menezes, and Gourab Ghoshal}                         

\date{}

\maketitle

\tableofcontents

\newpage

\section{Data}\label{sec:dataset}

\subsection{Dataset collection}\label{sec:data_collection}
\begin{itemize}
	\item{{\it BrightKite}} is a LBSN service provider that allowed registered users to connect with their existing social ties and also meet new people based on the places that they go. Once a user "checked in" at a place, they could post notes and photos to a location and other users could comment on those posts. The social relationship network was collected using their public API. Dataset link:  \href{https://snap.stanford.edu/data/loc-Brightkite.html}{https://snap.stanford.edu/data/loc-Brightkite.html}
	
	\item{{\it Gowalla}}
	This is a LBSN website where users share their locations by checking-in. In early versions of the service, users would occasionally receive a virtual "Item" as a bonus upon checking in, and these items could be swapped or dropped at other spots. Users became "Founders" of a spot by dropping an item there. 
	This incentivises users to create new check-ins, not necessarily to check-in consistently at frequently visited locations.  The social relationship network is undirected and was collected using their public API. Dataset link:  \href{https://snap.stanford.edu/data/loc-gowalla.html}{https://snap.stanford.edu/data/loc-gowalla.html}
	
	\item{{\it Weeplaces} - } 
	This is collected from Weeplaces and integrated with the APIs of other LBSN services, e.g., Facebook Places, Foursquare, and Gowalla. Users can login Weeplaces using their LBSN accounts and connect with their social ties in the same LBSN who have also used this application. Weeplaces visualizes your check-ins on a map. Unlike Gowalla, there is no direct incentive in Weeplaces to alter one's visitation habits or check-ins, so there should be a more accurate representation of a regular person's mobility patterns.
	Dataset link: \href{https://www.yongliu.org/datasets/}{https://www.yongliu.org/datasets/}
\end{itemize}

\begin{figure}[htbp]
	\centering
	\includegraphics[width=0.7\linewidth]{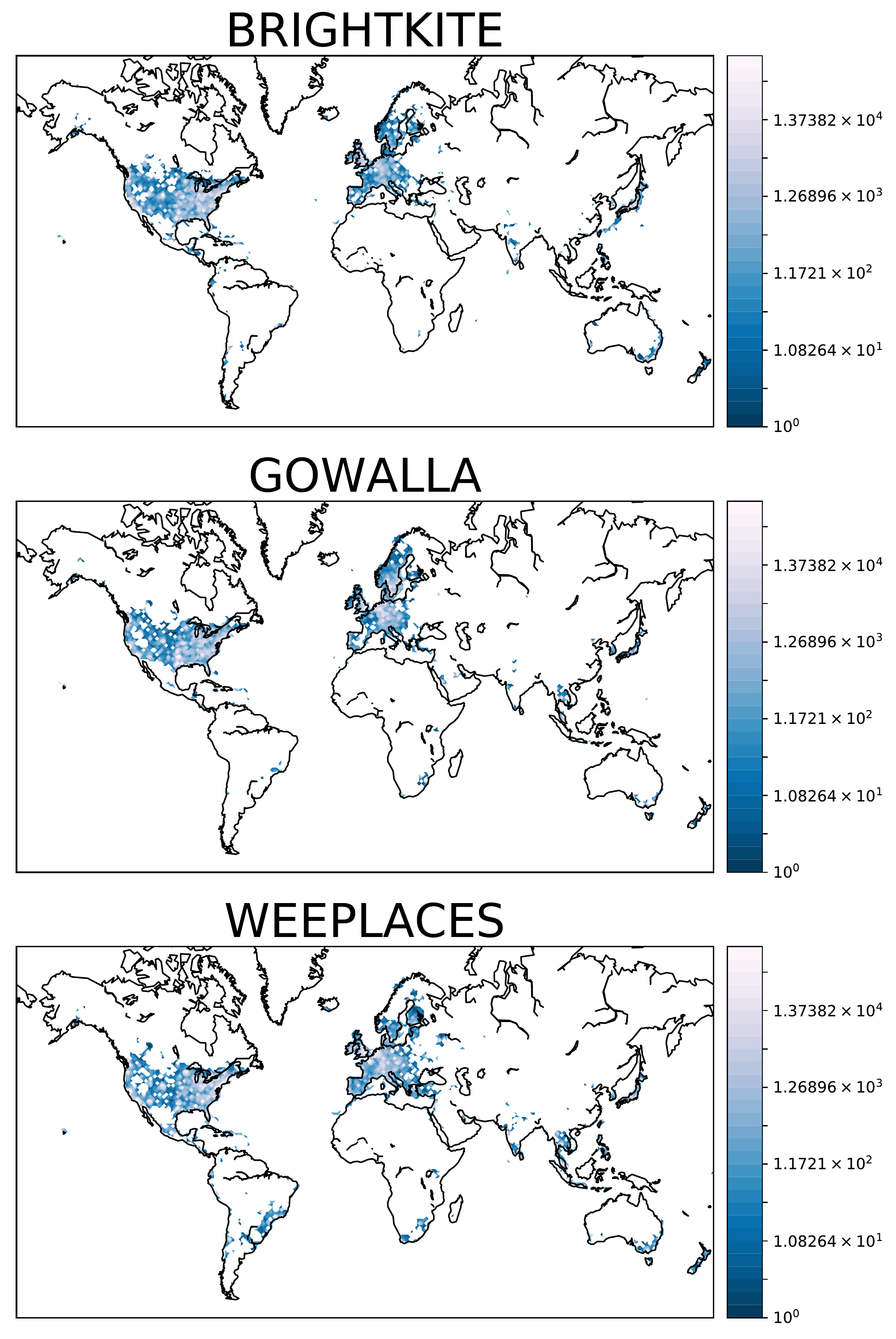}
	\caption{ \textbf{Check-in Maps of Gowalla, BrightKite, and Weeplaces.} The colorbar signifies the number of check-ins within a 50km radius, and shows highest coverage in North America and Western Europe}
	\label{fig:Check-in_Maps}
\end{figure}

\subsection{Mobility Statistics}
\begin{figure}[htbp]
	\centering
	\includegraphics[width=0.99\linewidth]{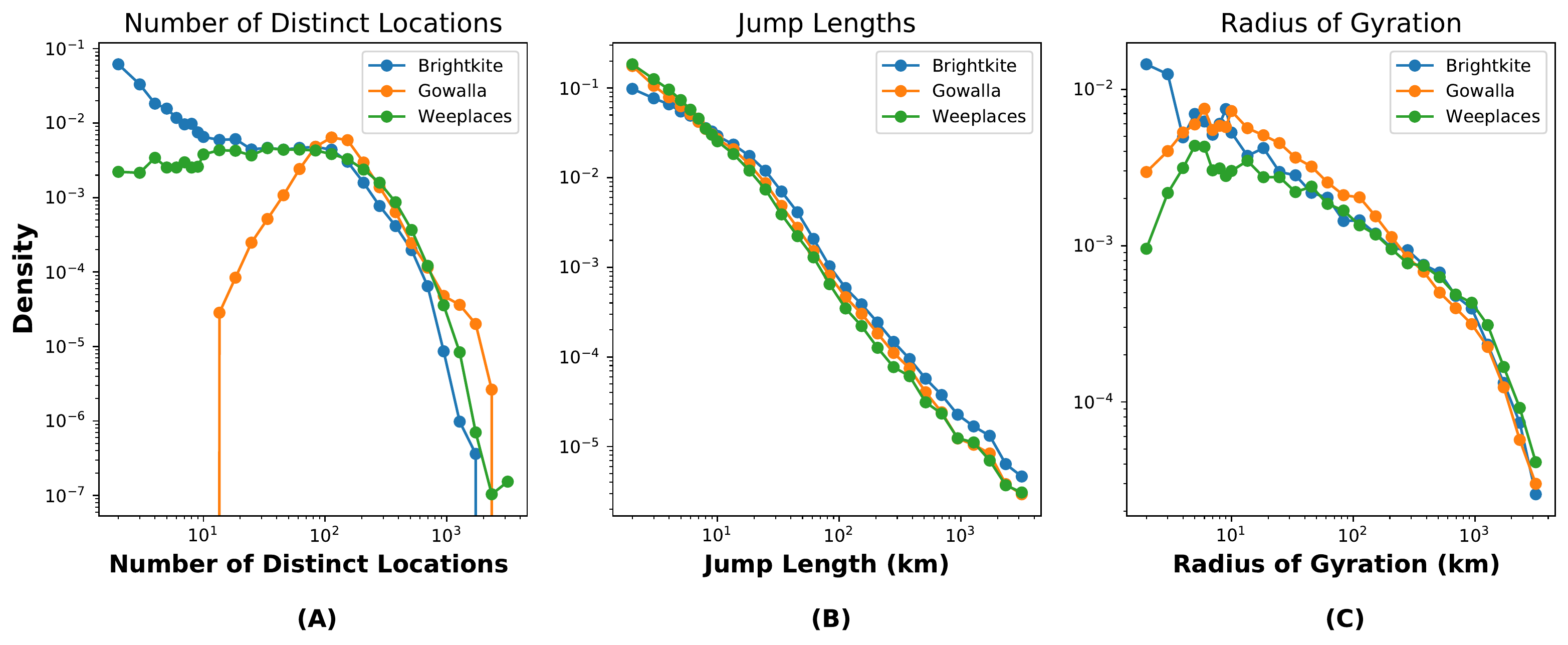}
	\caption{ \textbf{Statistical distributions of selected mobility quantities} \textbf{(A)} Distribution of total distinct locations visited by all users in each dataset. 
	\textbf{(B)} Distribution of Jump Lengths of all check-ins of users in the datasets
	\textbf{(C)} Histogram of Radii of Gyration for all users in the datasets
	}
	\label{fig:hist_DL_JL_RoG}
\end{figure}

 The number of unique visited locations, the distribution of jump lengths and the radii of gyration are shown in \autoref{fig:hist_DL_JL_RoG}. The latter two quantities are qualitatively identical among the three datasets and consistent with other sources. The distribution of jump lengths resembles a power law distribution, and the tail of the distribution of the radius of gyration closely represents a truncated power law. The distributions of the number of distinct locations illuminate characteristic differences between the datasets. Users of Gowalla are more likely to to visit many locations, while users of BrightKite check-in at very few distinct locations.

\subsection{Pre-processing}\label{sec:convergence}

\subsubsection{Entropy estimator convergence}
Because our data is finite and fairly sparse, we need to understand how well the entropy estimators saturate and set thresholds for our data for robustness. We establish a threshold for number of check-ins per user that will yield enough data to analyze the per-user entropy. We heuristically set a threshold of 150 check-ins, plot the entropy rate with respect to a percentage of the ego's trajectory, and find that the data roughly stabilizes within 50\% of the ego's trajectory.


\begin{figure}[htbp]
	\centering
    \includegraphics[width=0.8\linewidth]{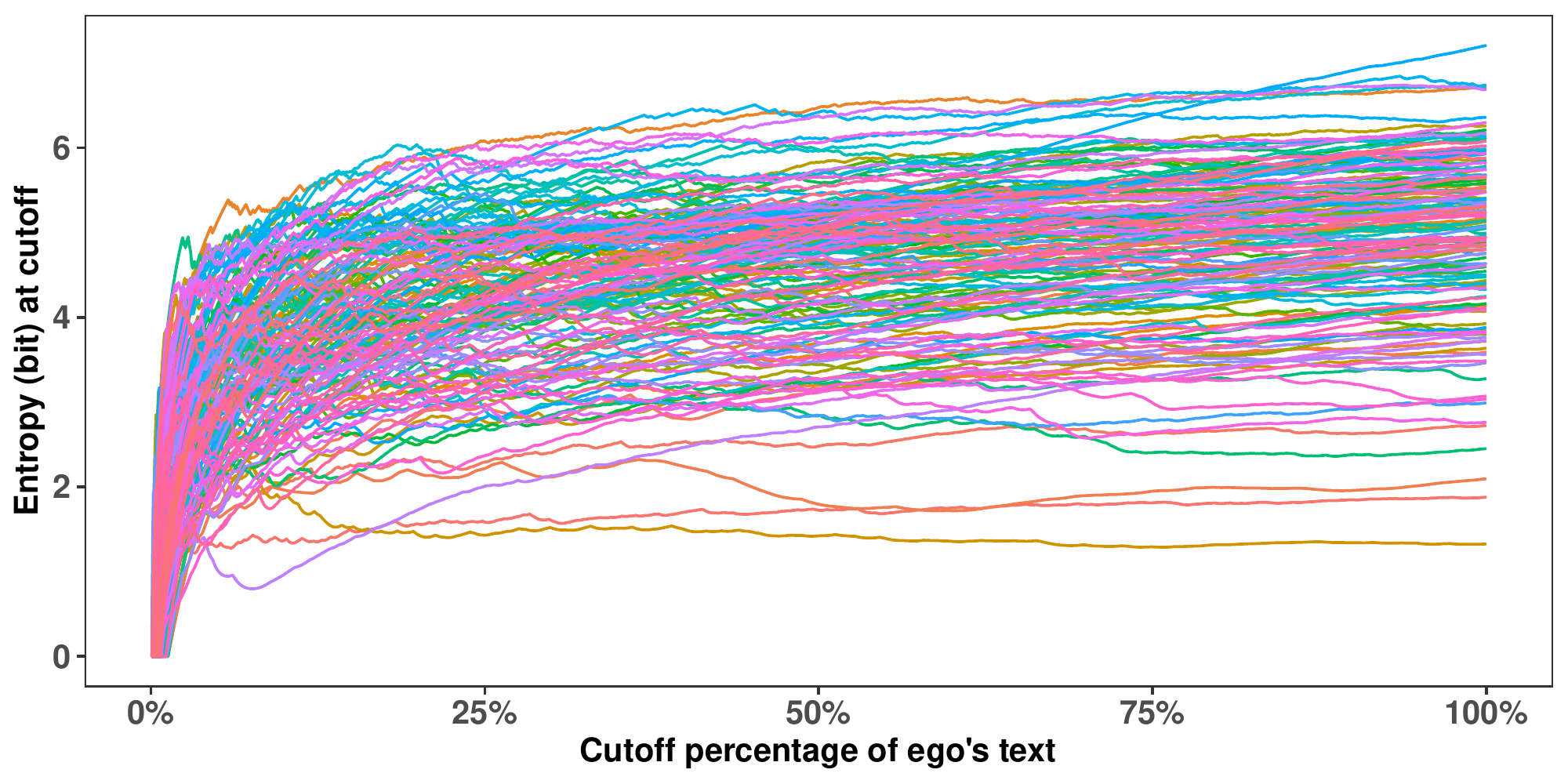}
	\caption{ \textbf{Entropy rate as a function of a percentage of the ego's trajectory.}  }
	\label{fig:LZ_entropy_convergence_network}
\end{figure}

\subsubsection{Cross-entropy estimator convergence}
\begin{figure}[htbp]
	\centering
    \includegraphics[width=0.98\linewidth]{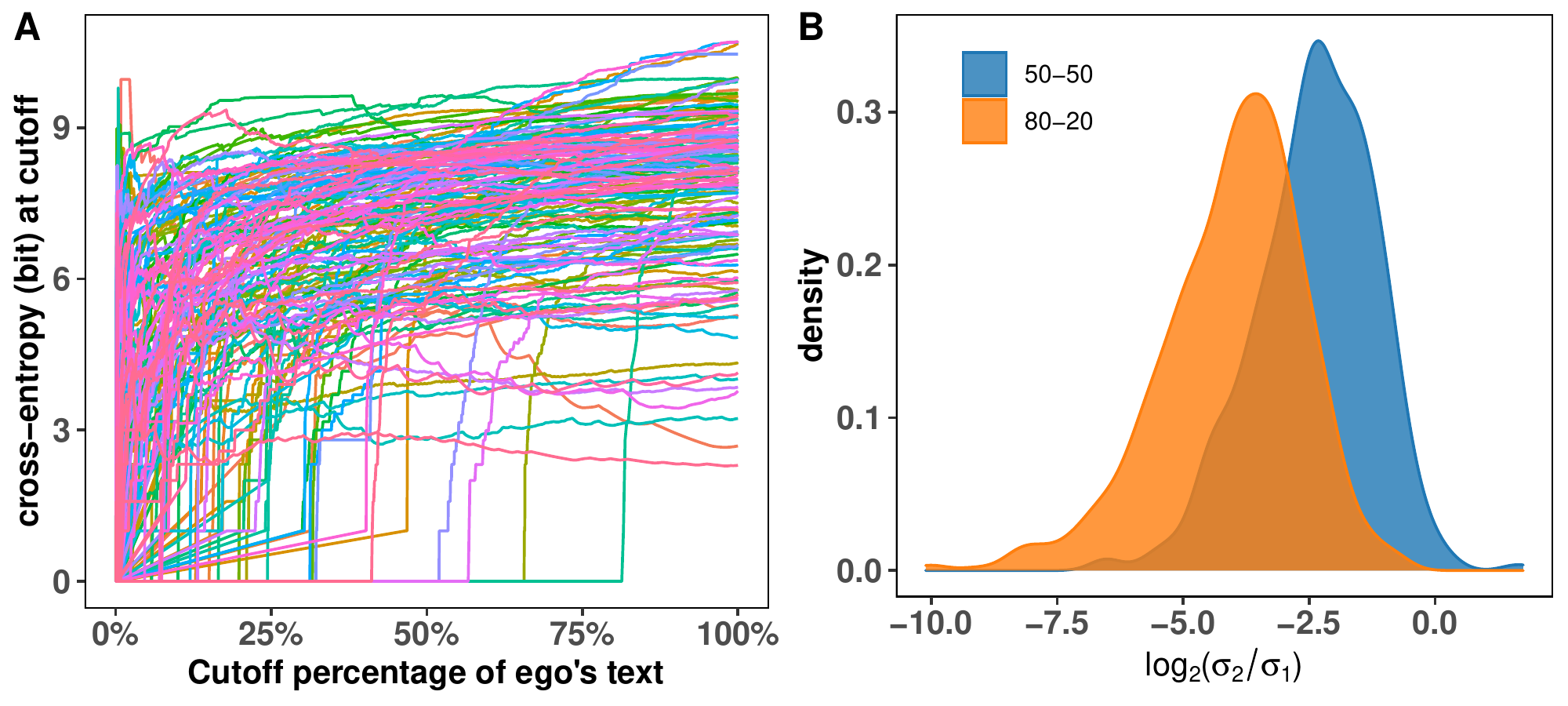}
	\caption{ \textbf{Convergence of Cross Entropy Estimator curve.} Shown are the cross entropies of the rank-1 alters for a subset of users (for ease of visualization). \textbf{(A)}, The cross entropy of rank-1 alters as a function of a cutoff percentage of the ego's trajectory
	\textbf{(B)}, The log ratio of the standard deviations of the end and beginning partitions show much lower variability in the latter end of the trajectory, signifying a leveling-off of the cross entropy estimator. }
	\label{fig:convergence_network}
\end{figure}

There are some pairs whose cross-entropy varies dramatically over the final portion of the data (\autoref{fig:convergence_network}A). To examine this we plot the cross entropy of rank-1 alters with respect to a cutoff\% of their ego's trajectory. First, we find many alters have check-ins after the last check-in of the ego, leading to a trailing sequence of check-ins that would over-inflate the entropy and do not contribute to the information contained in the ego. Therefore we establish a cutoff of \(N_{previous} = 150\), meaning there must be at least 150 alter-check-ins before the last check-in of the ego.



We also see from (\autoref{fig:convergence_network}A) that the distribution of check-ins among alters is varied, in that some alters do not have many check-ins at the beginning of the ego's trajectory, but they satisfy the \(N_{previous} >= 150\) threshold. To examine the saturation of the entropy and cross-entropy estimators, we partition the data into two portions and compare the variances of the entropy values of the two portions. We use a 50-50 split and an 80-20 split to show how well the cross entropy saturates in the final part of the data compared to the previous. The 80-20 split was chosen based on the 150 check-in minimum, so the final 20\% of the data would have at least 30 check-ins for which one could reasonably calculate a variance.
The two partitions show that as the number of check-ins increase, the variance in the latter portion relative to the earlier portion is much smaller, indicating that alters with high variability are few as seen in the rapid fall-off in the tail of the distribution.
After pre-processing, the summary statistics are shown in \autoref{tab:dataset}.

\begin{longtable}[c]{|c|c|c|c|}
\caption{\textbf{The summary of three pre-processed datasets}. 
}
\label{tab:dataset}\\
\hline
\multicolumn{1}{|c|}{Dataset} & Total   Check-ins & Users  & Distinct   Placeid \\ \hline
\endfirsthead
\multicolumn{4}{c}%
{{\bfseries Table \thetable\ continued from previous page}} \\
\endhead
Weeplace                      & 7,049,037         & 11,533 & 924,666            \\ \hline
BrightKite                    & 3,513,895         & 6,132  & 510,308            \\ \hline
Gowalla                       & 3,466,392         & 9,937  & 850,094            \\ \hline
\end{longtable}

\section{Ego-alter network construction}\label{sec:network-construction}

In each of the LBSNs we use, there exist both a location check-in network and a social network, so we can use the social network to compare against a proxy network. With the check-in network, we can form an artificial social network by assigning connections to users that check in at the same place at the same time. We took a colocation as two users checking in at the same place-id within an hour starting on the hour, eg. 8:00 - 9:00 (The choice of the 1-HR bin for colocation was examined and the details can be found in \autoref{sec:robustness}). We assume that users that co-locate more often contain more predictive information about each other's whereabouts, so we rank users' social relationship based on the number of times they co-locate, both in the social relationship and colocation network. Because two people that co-locate are not necessarily social ties, we use the term "ego" to describe users who's mobility data we are trying to predict by the location history of their "alters" (non-social colocators and social ties).

\subsection{Quality control of colocators}\label{sec:quality_control}
Unlike social relationship, colocation is a theoretical and artificial relationship between two individuals. We find that a user among all datasets reasonably have more colocators than social ties. There are many colocators who are strangers to the user because the two happen to check-in at the same time by accident, eg. a colocator with a single colocation with a user. 
If user $A$ and $B$ only have one colocation, they are probably strangers to each other and just happen to appear in one place at the same time period by chance. Thus, there is no doubt that $A$ (or $B$) can hardly provide any useful information for $B$ (or $A$) when looking at their previous trajectories.
Therefore, we made further \emph{quality control} by discarding the colocator with only one colocation.

For all users in the datasets, their remaining colocators make up the \emph{qualified} network and these colocators are called \emph{qualified} colocators. All colocators used in this paper are qualified colocators, unless indicated otherwise.

\subsection{Contribution control of colocators and social ties}\label{sec:contribution_control}
We can use a random algorithm as null model to compare to the information contained in a user's trajectory. The probability of correct prediction in null model is $1/N_{ego}$ where $N_{ego}$ is the number of unique locations of ego's historical trajectory.
Any useful algorithms have to perform better than a random algorithm and these algorithms should have no fewer than $\log_2(N_{ego})$ bits information from the view of information theory. As a consequence, we took further \emph{contribution control} of both colocators and social ties according to LZ cross-entropy. For any ego, we classified all alters (colocators and social ties) as two groups: "\emph{useful}" if the ego-alter pair's LZ cross-entropy is less than $\log_2(N_{ego})$, and  "\emph{useless}" otherwise. Additional colocators were found whose sequences had no previous matches to their ego's sequence at any point in the ego's trajectory (\(w_b = 0\) in \autoref{eq:CCE}). We therefore consider all pairs where the alter's trajectory contained no previous matches as "useless". After these considerations were made, we considered only the \textit{useful} alters and discarded the rest.


\subsection{Statistics of constructed network}

\begin{longtable}[c]{|c|c|c|c|c|c|c|}
\caption{\textbf{The summary of three filtered datasets}. The number of egos who have at least ten alters in both non-social colocation (quality control and contribution control apply) and social (contribution control applies) network. The common networks include the common egos with their respective top ten alters in each network.}
\label{tab:CLN_SRN_stats}\\
\hline
\multicolumn{1}{|c|}{\multirow{2}{*}{Dataset}} &
  \multicolumn{2}{c|}{Non-social colocation network} &
  \multicolumn{2}{c|}{Social network} &
  \multicolumn{2}{c|}{Common-Ego networks} \\ \cline{2-7} 
\multicolumn{1}{|c|}{} & ego & ego-alter pair & ego & ego-alter pair & ego & ego-alter pair \\ \hline
\endfirsthead
\multicolumn{7}{c}%
{{\bfseries Table \thetable\ continued from previous page}} \\
\endhead
BrightKite             & 122 & 2,684           & 187 & 4,460           & 33  & 330            \\ \hline
Gowalla                & 192 & 9,332          & 349 & 7,681           & 97 & 970           \\ \hline
Weeplaces              & 665 & 21,741          & 401 & 8,042           & 199 & 1,990           \\ \hline
\end{longtable}


\subsection{Choice of the number of top alters}\label{sec:choice_number_top_alters}

\begin{figure}[htbp]
	\centering
	\includegraphics[width=0.7\linewidth]{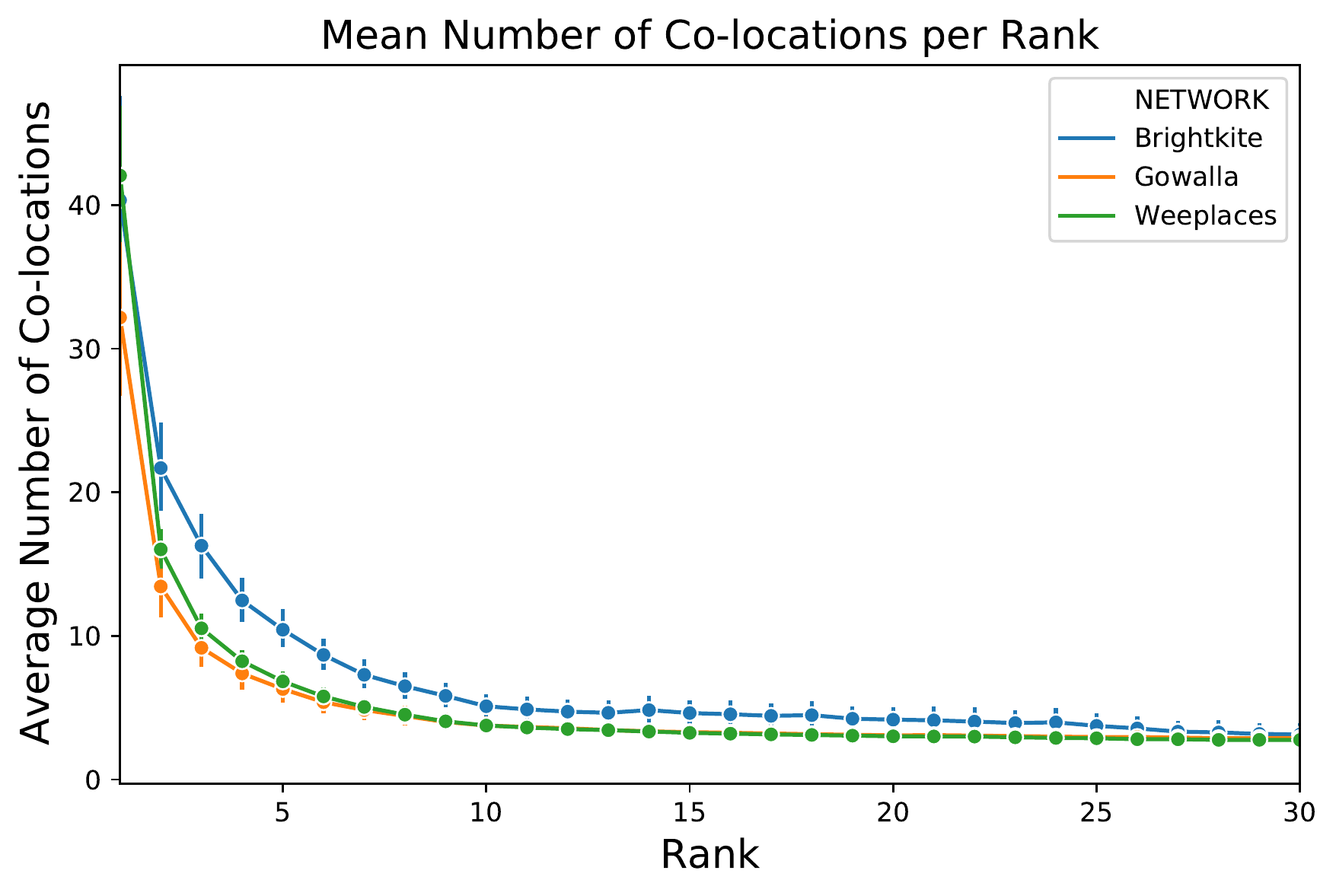}
	\caption{ \textbf{Average number of colocations per rank for all datasets}}
	\label{fig:convergence_ratio}
\end{figure}

\begin{figure}[htbp]
	\centering
	\includegraphics[width=0.82\linewidth]{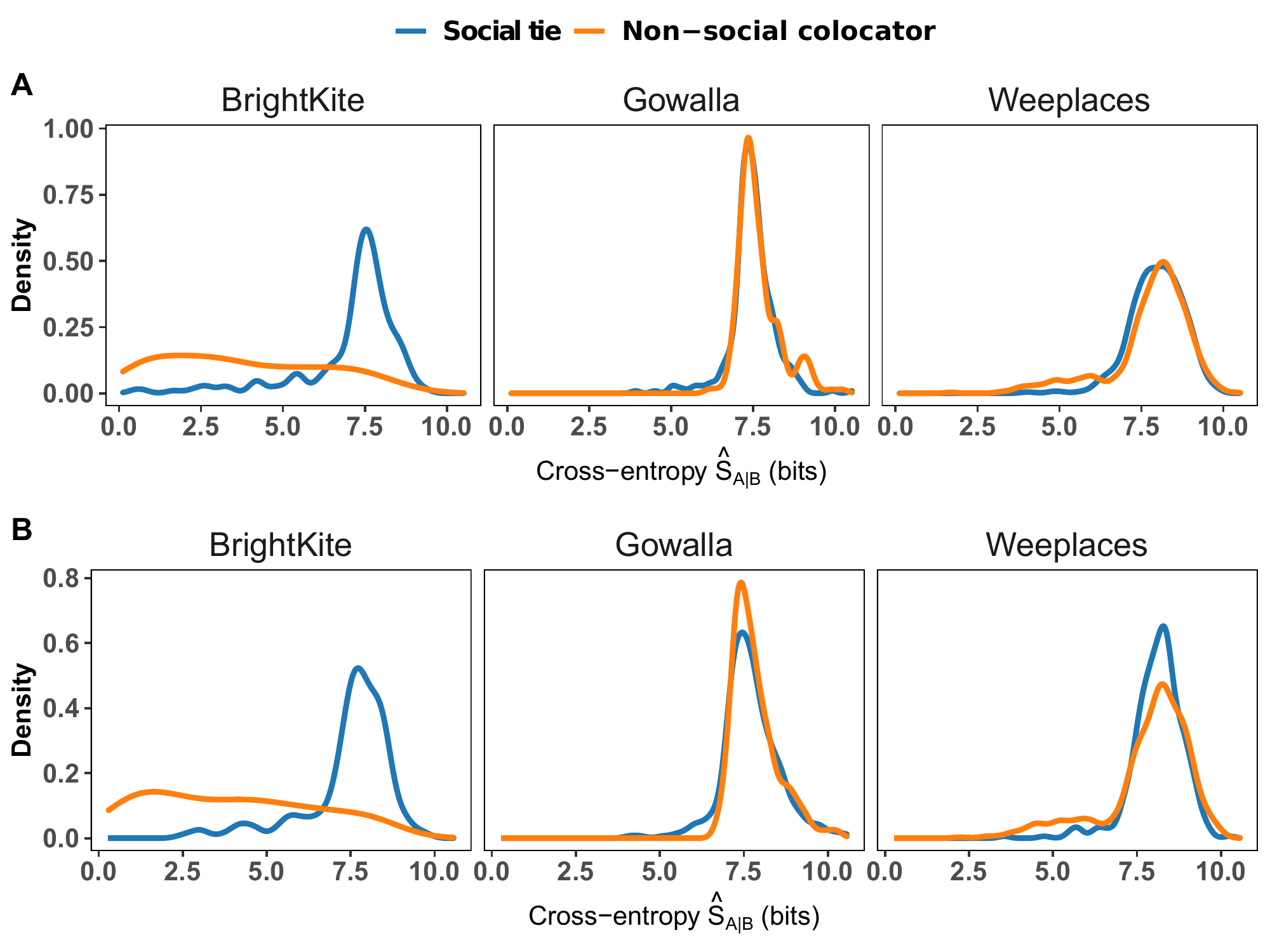}
	\caption{ \textbf{Information contained in alters.} \textbf{(A)} Cross-entropy of ego with their rank-5 social tie and rank-5 non-social colocator. 
	\textbf{(B)} Cross-entropy of ego with their rank-10 social tie and rank-10 non-social colocator. 
	}
	\label{fig:hist_CE_CLN_SRN}
\end{figure}

We would like to choose the highest quality subset of alters among our data. Because we choose colocation as a proxy for social ties, we examine the usefulness of alters based on the number of colocations. We place an additional sub-ranking criteria where alters with the same number of colocations are ranked in increasing order of number of check-ins of the alter. We assume that alters with higher ratios of colocations to number of checkins will provide more information than those with lower ratios. After keeping all useful alters, we look at the average number of colocations per rank of the datasets (\autoref{fig:convergence_ratio}). All three datasets show that beyond roughly 10 alters, the number of colocations saturates at fewer than around 5 meetups, making the potential influence due to colocation insignificant. Therefore, in our analysis we focused on the contribution from the top ten alters. Cross Entropy distributions of rank-5 (middle) and rank-10 (low) were plotted for both networks for all datasets, and on average the entropies increased with lower rank (\autoref{fig:hist_CE_CLN_SRN}).

\subsection{Robustness of temporal windows in colocation network construction }\label{sec:robustness}

\begin{figure}[htbp]
	\centering
	\includegraphics[width=0.80\linewidth]{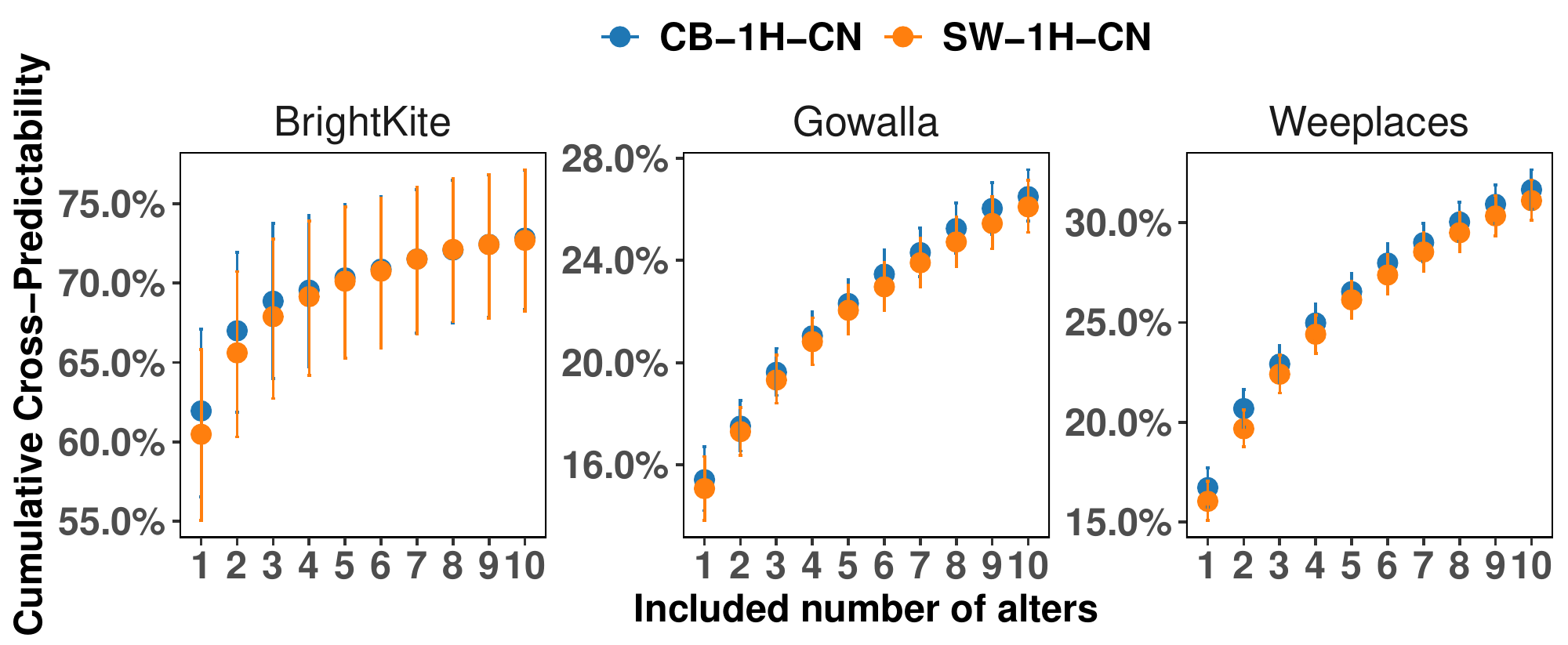}
	\caption{ \textbf{The comparison between choice of 1-hour clock-bin and sliding window in BrightKite, Gowalla, and Weeplaces}.
    Error bars denote $\pm95\%$ CI.
	}
	\label{fig:wp_CB-1h_vsSW-1h}
\end{figure}

The colocation network construction relies on a specified time-resolution, so different time-resolutions would capture different colocators for each ego. To test the robustness of the 1-hour colocation time frame, we compared a 1-hour clock-bin network (colocation on a given day within the interval ($T$:00:00,$T$:59:59), $T\in (0,1,2...,23)$) to a 1-hour sliding-window network (colocation within $\pm 30$ minutes of a check-in of the ego). 
We consider the 199 egos with at least ten alters in Weeplaces dataset. From \autoref{fig:wp_CB-1h_vsSW-1h} we see that within error bars, the colocation networks of different-sized clock bins have statistically similar trends in cumulative cross-predictability. 
The cumulative predictabilities of ego with increasing number of qualified colocators based on different temporal windows are statistically similar.

\section{Information contained in alters}\label{sec:CE-All-Datasets}

\subsection{Analysis for Brightkite and Gowalla}

\begin{figure}[htbp!]
	\centering
	\includegraphics[width=0.99\linewidth]{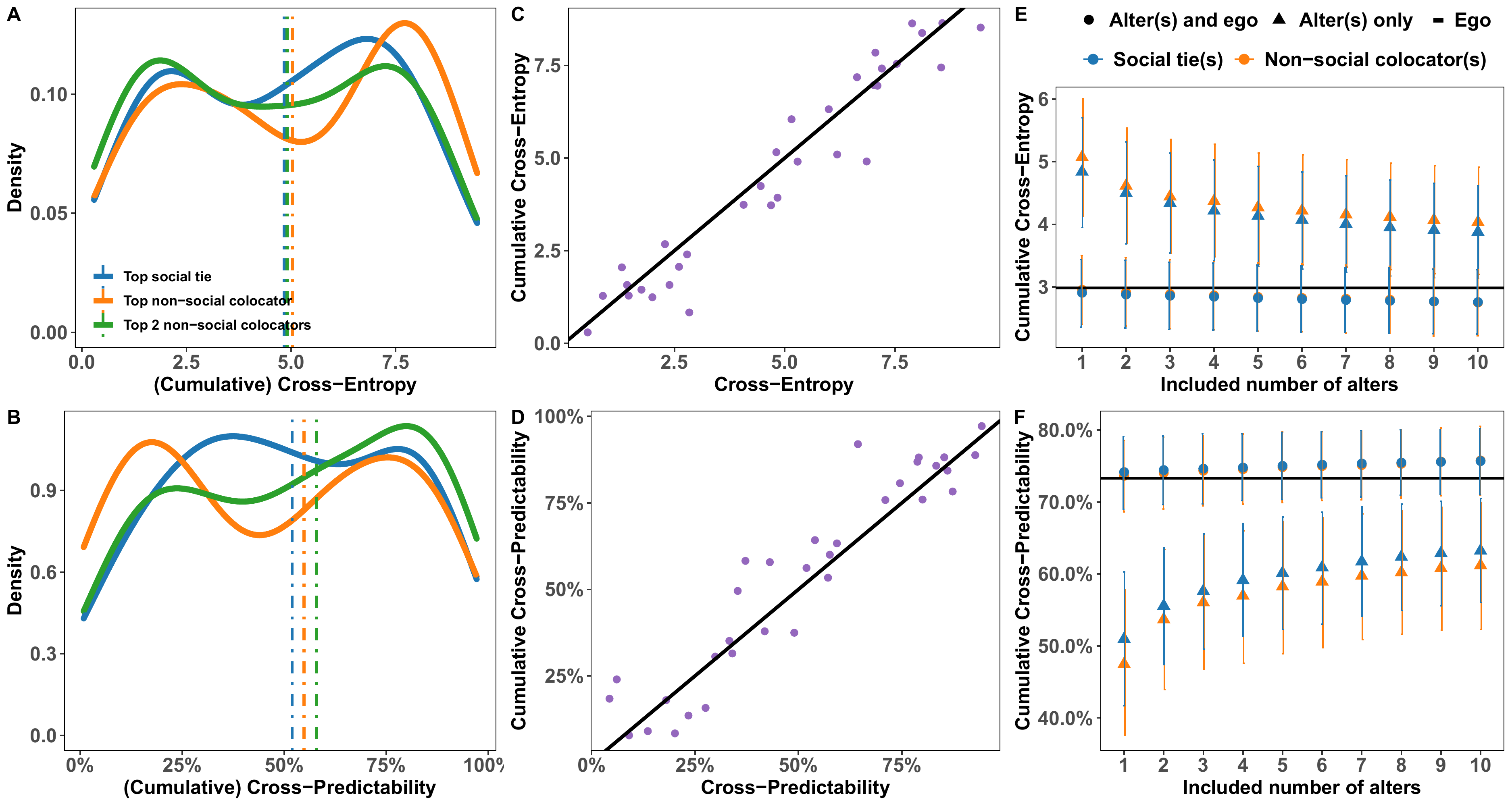}
	\caption{ \textbf{The cross-entropy and predictability provided by social ties and non-social colocators in BrightKite}.
	\textbf{A} 
	Distributions of $\hat S_{A|B}$ for the rank-1 social tie (median 4.84 bits), non-social colocator (median 5.03 bits), and $\hat S_{A|\mathcal{B}}$ for the top-2 non-social colocators (median 4.90 bits) in 
	\textbf{B}
	The corresponding $\Pi_{A|B}$ for the social  (median 51.94\%), and non-social colocators (median 54.84\%), and  $\Pi_{A|\mathcal{B}}$ for the top-2 non-social colocators (median 57.86\%).  
	\textbf{C}
	$\hat S_{A|B}$ encoded in the top-social tie as a function of $\hat S_{A|\mathcal{B}}$ for the top-3 non-social colocators. Each point corresponds to a single ego and the solid line denotes $y = x$.
	\textbf{D}
	As in panel {\bf C} but with predictability instead of cross-entropy.
	\textbf{E, F} 
	$\hat S_{A|\mathcal{B}}$ and $\Pi_{A|\mathcal{B}}$ after accumulating the top-ten social alters and non-social colocators. 
	Horizontal lines denote the average entropy ($2.98$ bits) of egos and their self-predictability ($73.33$\%).
	Error bars denote $95\%$ CI. 
	}
	\label{fig:bk_CCP_social_vs_non_social}
\end{figure}

\begin{figure}[htbp!]
	\centering
	\includegraphics[width=0.99\linewidth]{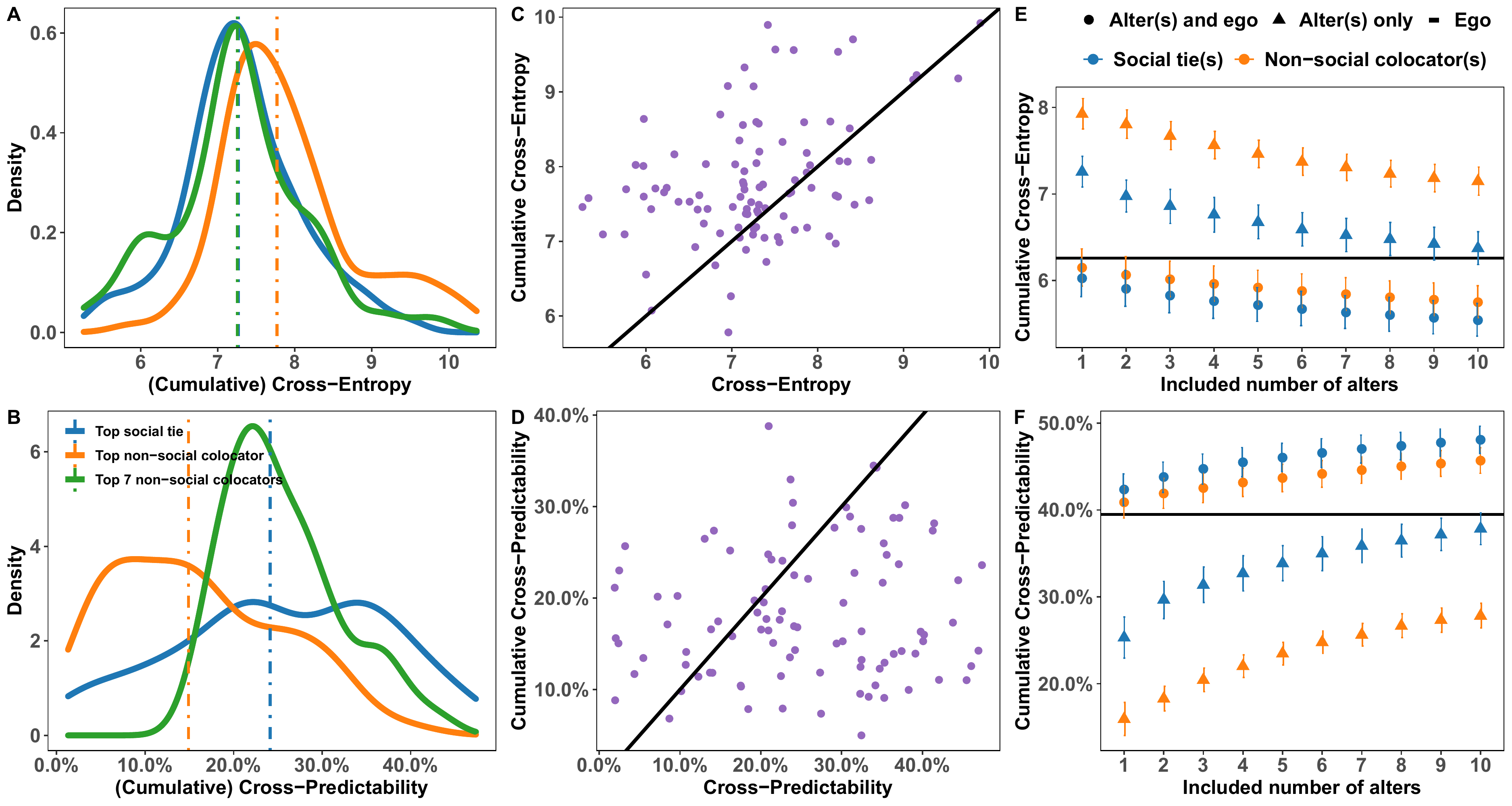}
	
	\caption{ \textbf{The cross-entropy and predictability provided by social ties and non-social colocators in Gowalla}.
	\textbf{A} 
	Distributions of $\hat S_{A|B}$ for the rank-1 social tie (median 7.27 bits), non-social colocator (median 7.77 bits), and $\hat S_{A|\mathcal{B}}$ for the top-7 non-social colocators (median 7.26 bits) in 
	\textbf{B}
	The corresponding $\Pi_{A|B}$ for the social  (median 24.14\%), and non-social colocators (median 14.92\%), and  $\Pi_{A|\mathcal{B}}$ for the top-7 non-social colocators (median 24.14\%).  
	\textbf{C}
	$\hat S_{A|B}$ encoded in the top-social tie as a function of $\hat S_{A|\mathcal{B}}$ for the top-3 non-social colocators. Each point corresponds to a single ego and the solid line denotes $y = x$.
	\textbf{D}
	As in panel {\bf C} but with predictability instead of cross-entropy.
	\textbf{E, F} 
	$\hat S_{A|\mathcal{B}}$ and $\Pi_{A|\mathcal{B}}$ after accumulating the top-ten social alters and non-social colocators. 
	Horizontal lines denote the average entropy ($6.26$ bits) of egos and their self-predictability ($39.49$\%).
	Error bars denote $95\%$ CI.
	}
	\label{fig:gw_CCP_social_vs_non_social}
\end{figure}

\autoref{fig:bk_CCP_social_vs_non_social} and \autoref{fig:gw_CCP_social_vs_non_social} show the information-theoretic analysis of Brightkite and Gowalla respectively

\begin{figure}[htbp]
	\centering
	\includegraphics[width=0.98\linewidth]{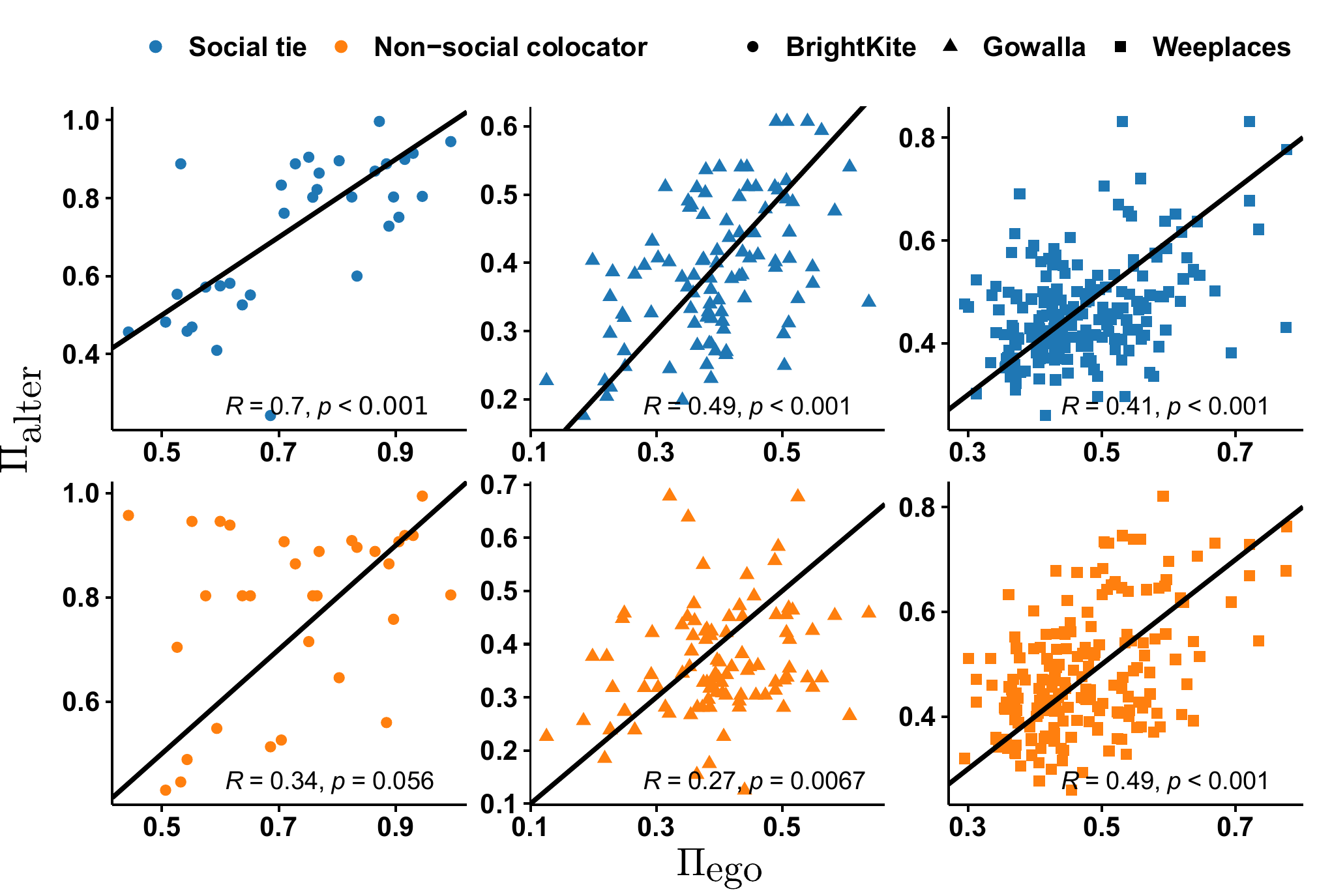}
	\caption{ 
	\textbf{Homophily in predictability}.
    Scatterplot comparing the predictabilities of egos to their rank-1 alters. All egos are those who have at least both 10 social ties and 10 non-social colocators. The black solid lines in each subplot are $y=x$.
	}
	\label{fig:homophily_analysis}
\end{figure}

\clearpage
\section{Extrapolating Cross-Predictability}\label{sec:saturate_fit}
We've chosen the top 10 alters in determining the cumulative mobility information flow between the alters' respective egos. We extrapolate these results by fitting a saturating function to our data, to determine the potential information flow in the limit of infinite alters (or more realistically around 150 alters, the maximum number of social ties a given person can reasonably have). The saturating function used is
\begin{equation}
    \Pi(i) = \Pi_{\infty} + \frac{\beta_0}{\beta_1 + i}
\end{equation}
where $i$ is the number of top $i$ included alters. A $\chi^2$ minimization of the means and their errors using the BFGS algorithm was used to determine the most likely parameters. A $95\%$  confidence interval of the parameters was determined using a t-test with $10$ alters $- 3$  parameters $= 7$ degrees of freedom. Results can be found in \autoref{tab:CLN_SRN_CCP_Fits}.

\begin{longtable}[c]{|p{2cm}|p{1.7cm}|p{2.8cm}|p{1.7cm}|p{2.8cm}|p{1.7cm}|p{2.8cm}|}
\caption{\textbf{Parameters for saturating function of the cumulative cross predictability $\Pi(i)$}}
\label{tab:CLN_SRN_CCP_Fits}\\
\hline
\multicolumn{1}{|c|}{\multirow{2}{2cm}{Dataset \&}} &
  \multicolumn{2}{c|}{Brightkite} &
  \multicolumn{2}{c|}{Gowalla} &
  \multicolumn{2}{c|}{Weeplaces} \\ \cline{2-7} 
\multicolumn{1}{|c|}{Network} & Social & Non-Social colocation & Social & Non-social colocation & Social & Non-social colocation \\ \hline
\endfirsthead
\multicolumn{7}{c}%
{{\bfseries Table \thetable\ continued from previous page}} \\
\endhead
$\Pi_{\infty}$         & 0.6699 $\pm$ .003313 & 0.6329 $\pm$ .00504 & .4319 $\pm$ 0.005082 & .3897 $\pm$ .003186  & 0.4431 $\pm$ .001039  &  0.3979 $\pm$  0.003427       
\\ \hline
$\beta_{0}$         & -0.4629 $\pm$ .03817 & -0.2980 $\pm$ .0445  & -.7427 $\pm$ .07489 & -1.881 $\pm$ .0747    & -0.7792 $\pm$ 0.01240 & -1.616 $\pm$ .0662           \\ \hline
$\beta_{1}$         & 1.937 $\pm$ .2011 & .9096 $\pm$ .25089          & 3.224 $\pm$ .3323 & 7.135 $\pm$ .2237           & 2.083 $\pm$ 0.04045 & 5.307 $\pm$ .1865           \\ \hline
\end{longtable}


%

\section{Spatial correlation analysis}    
\label{sec:correlationanalysis}

\begin{figure}[htbp]
	\centering
	\includegraphics[width=0.99\linewidth]{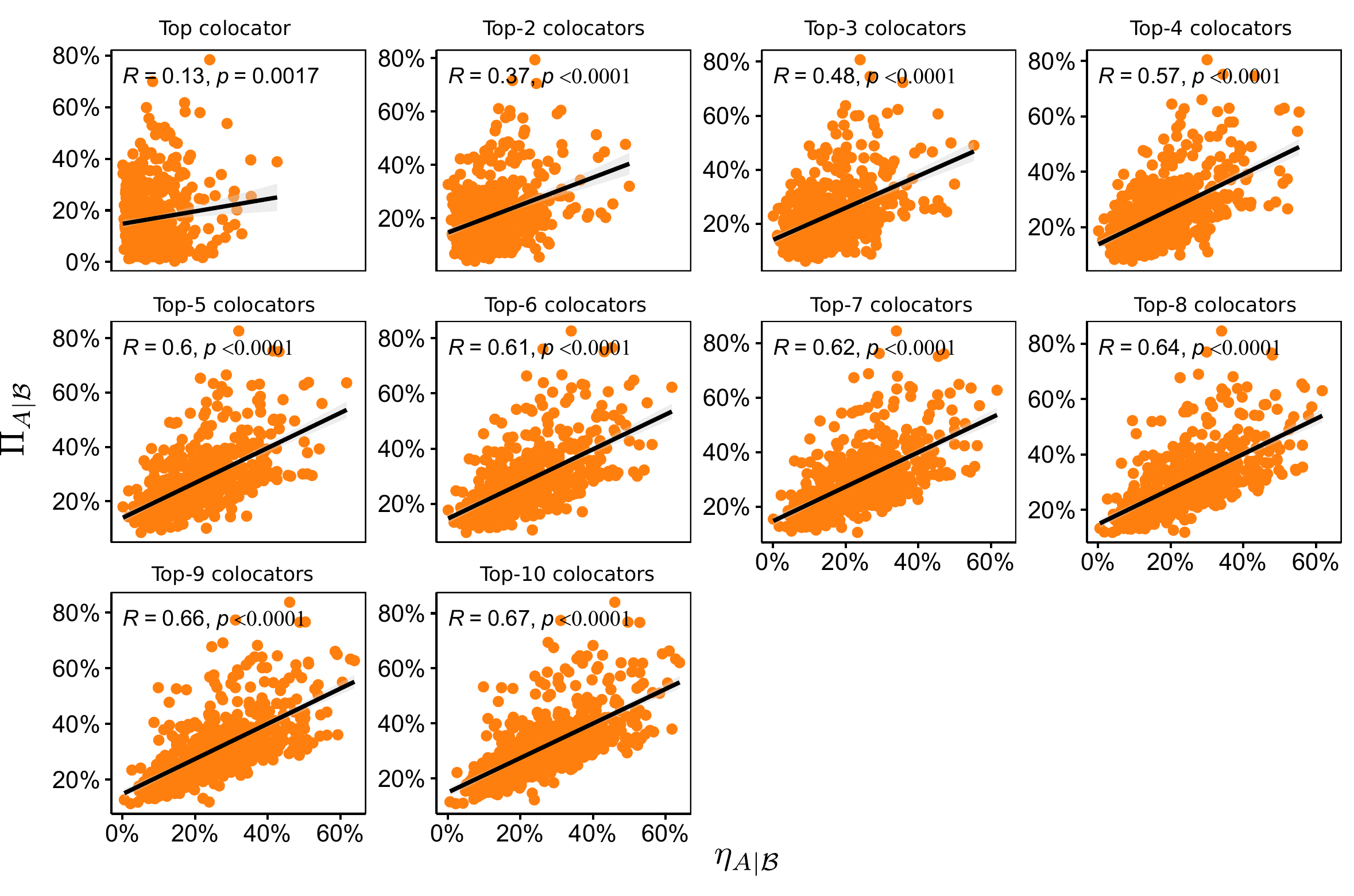}
	\caption{ \textbf{CODLR vs cumulative cross-predictability for non-social ties in Weeplaces.}
    $R$ is Pearson's correlation coefficient and $p$ is p-value. The solid black lines are linear regression lines.}
	\label{fig:vip_MeetupNp_CODLR_CCP_wp_H_MFN}
\end{figure}

\begin{figure}[htbp]
	\centering
	\includegraphics[width=0.99\linewidth]{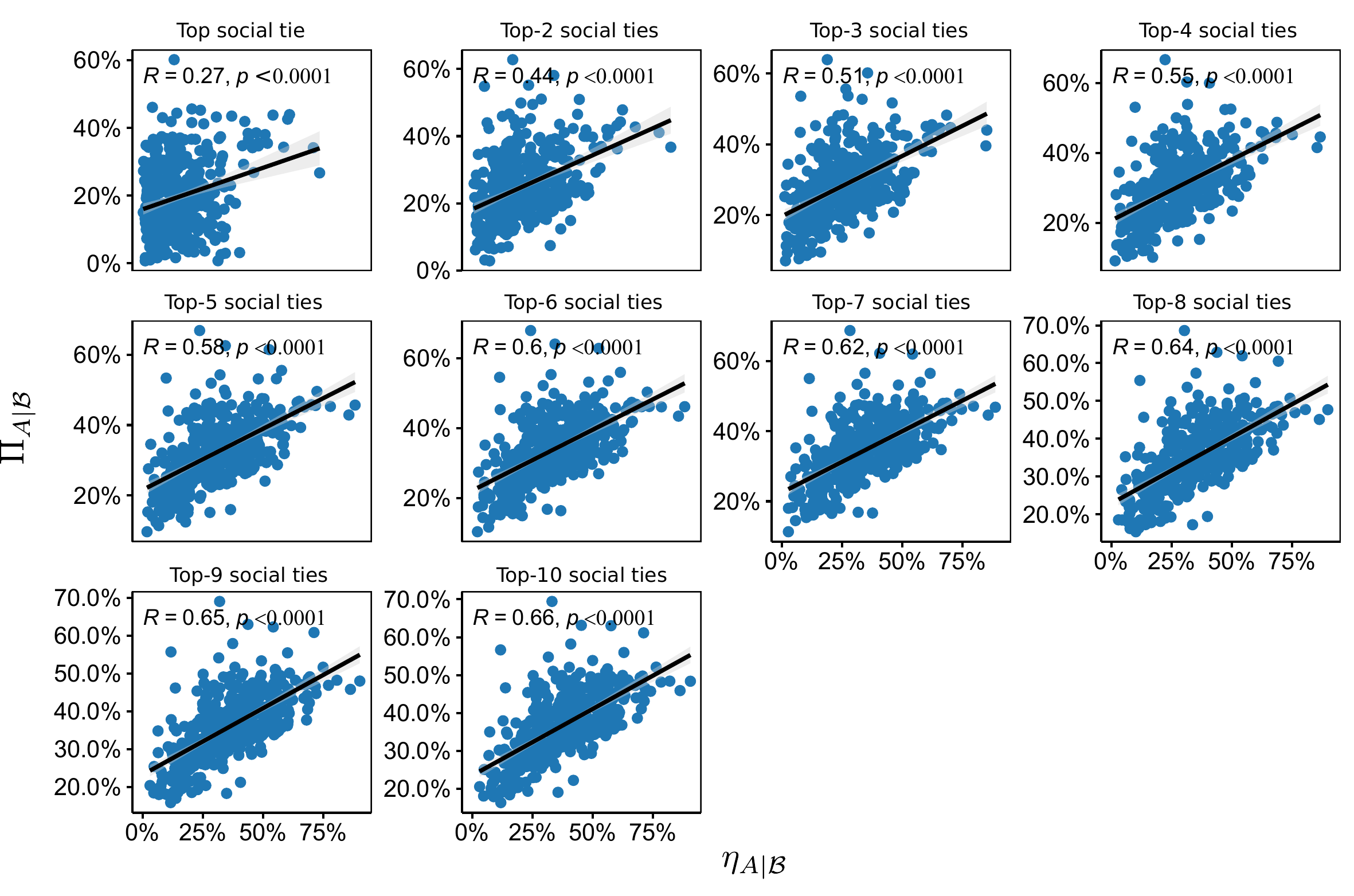}
	\caption{ \textbf{CODLR vs cumulative cross-predictability for social ties in Weeplaces.}
	$R$ is Pearson's correlation coefficient and $p$ is p-value. The solid black lines are linear regression lines.}
	\label{fig:vip_MeetupNp_CODLR_CCP_wp_TFN}
\end{figure}

\begin{figure}[htbp]
	\centering
	\includegraphics[width=0.99\linewidth]{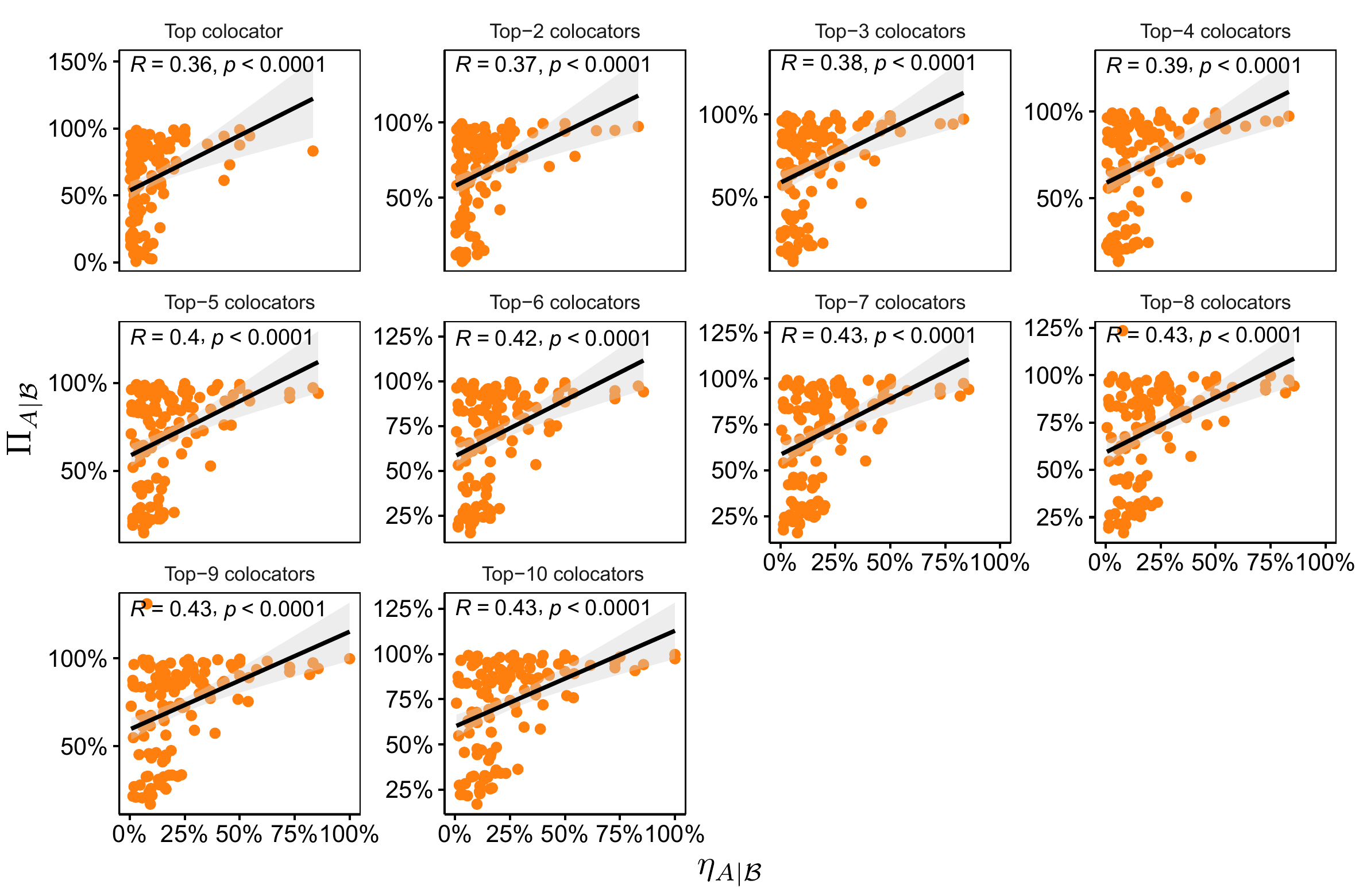}
	\caption{ \textbf{CODLR vs cumulative cross-predictability for non-social ties in BrightKite.}
    $R$ is Pearson's correlation coefficient and $p$ is p-value. The solid black lines are linear regression lines.}
	\label{fig:vip_MeetupNp_CODLR_CCP_bk_H_MFN}
\end{figure}

\begin{figure}[htbp]
	\centering
	\includegraphics[width=0.99\linewidth]{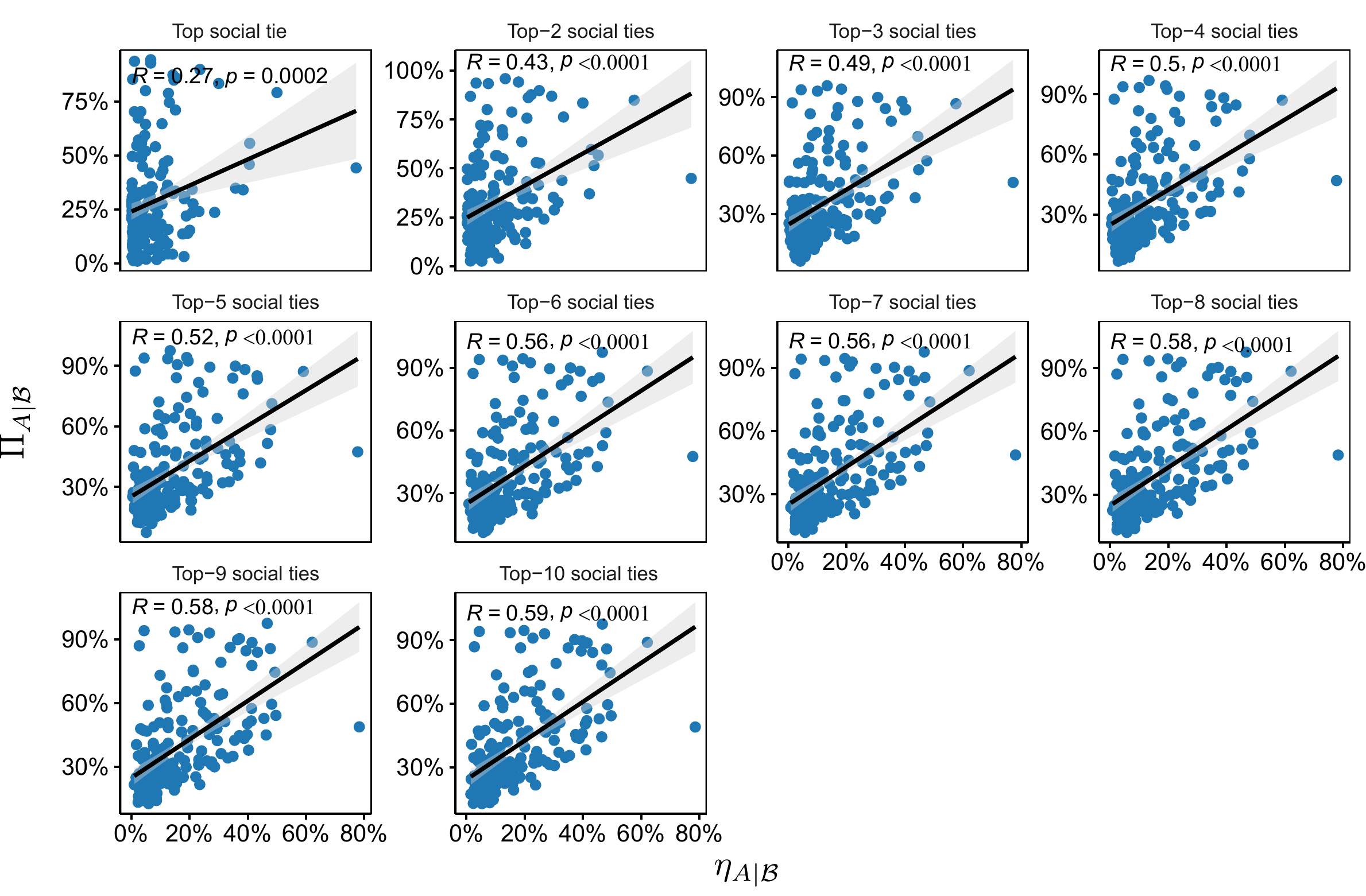}
	\caption{ \textbf{CODLR vs cumulative cross-predictability for social ties in BrightKite.}
	$R$ is Pearson's correlation coefficient and $p$ is p-value. The solid black lines are linear regression lines.}
	\label{fig:vip_MeetupNp_CODLR_CCP_bk_TFN}
\end{figure}

\begin{figure}[htbp]
	\centering
	\includegraphics[width=0.99\linewidth]{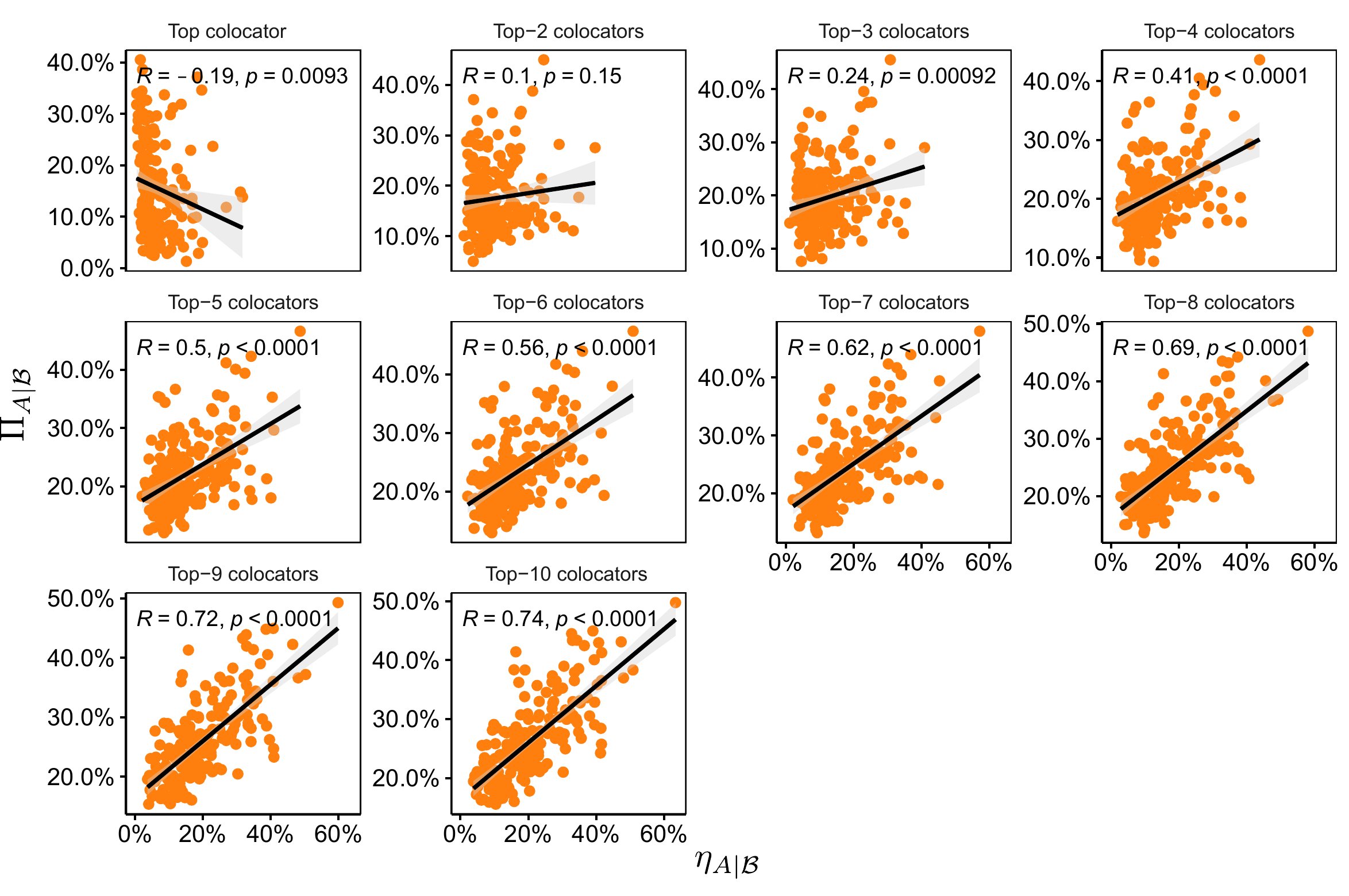}
	\caption{ \textbf{CODLR vs cumulative cross-predictability for non-social ties in Gowalla.}
    $R$ is Pearson's correlation coefficient and $p$ is p-value. The solid black lines are linear regression lines.}
	\label{fig:vip_MeetupNp_CODLR_CCP_gw_H_MFN}
\end{figure}

\begin{figure}[htbp]
	\centering
	\includegraphics[width=0.99\linewidth]{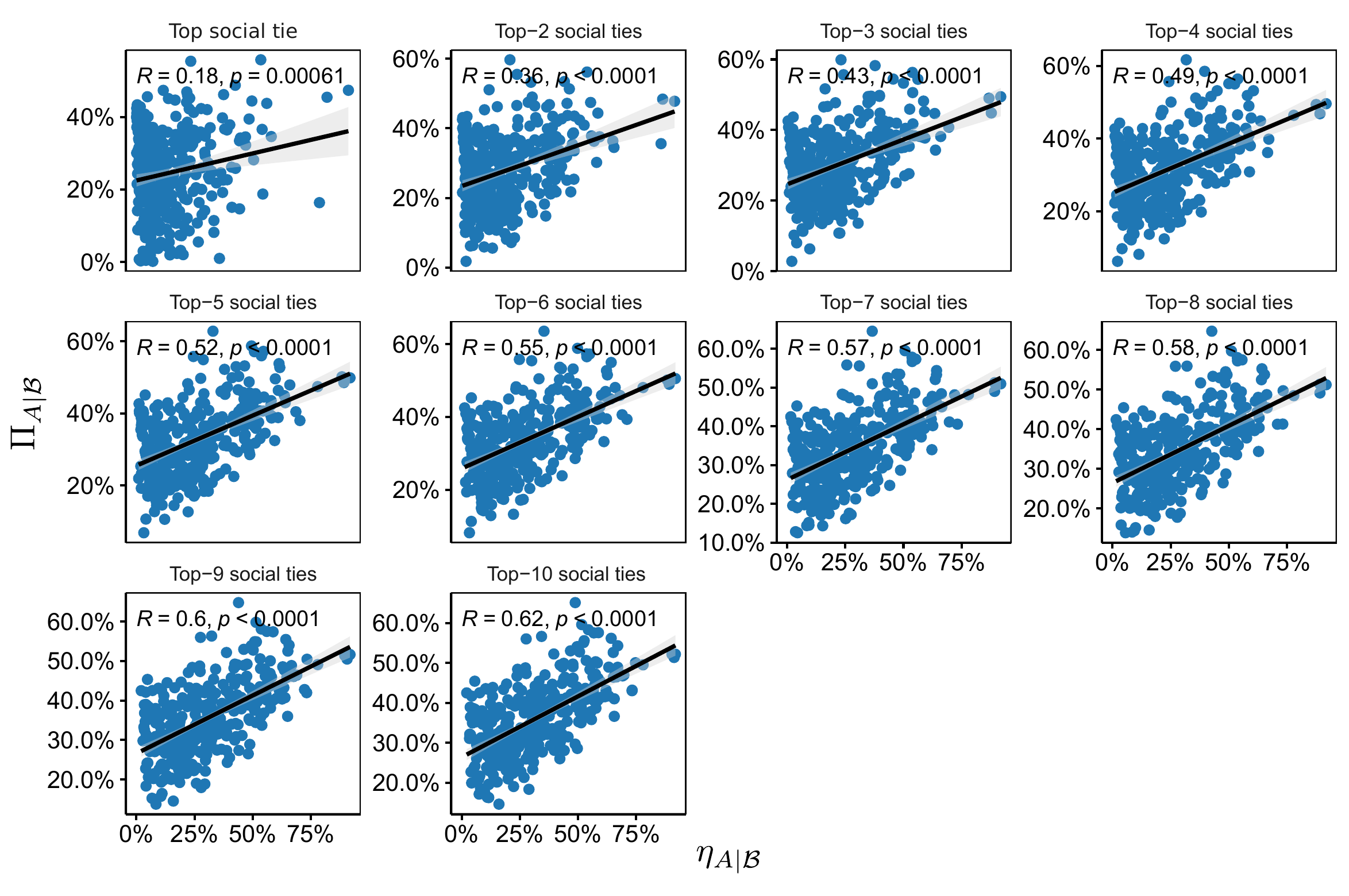}
	\caption{ \textbf{CODLR vs cumulative cross-predictability for social ties  in Gowalla.}
	$R$ is Pearson's correlation coefficient and $p$ is p-value. The solid black lines are linear regression lines.}
	\label{fig:vip_MeetupNp_CODLR_CCP_gw_TFN}
\end{figure}

We plot the correlation between the cumulative cross-predictability and the CODLR in both types of networks as one progressively adds alters from rank-1 to rank -10 in \autoref{fig:vip_MeetupNp_CODLR_CCP_wp_H_MFN} and \autoref{fig:vip_MeetupNp_CODLR_CCP_wp_TFN} for the Weeplaces dataset (571 common egos). While including a single alter yields a Pearson correlation coefficient $R = 0.13$ in colocation network and $R = 0.27$ in social network, the correlation increases as one progressively adds more alters saturating at $R=0.67$ and $R=0.66$ in colocation network and social network,respectively.
We can also the same trend in both BrightKite (See \autoref{fig:vip_MeetupNp_CODLR_CCP_bk_H_MFN} and \autoref{fig:vip_MeetupNp_CODLR_CCP_bk_TFN}, 122 common egos)and Gowalla ((See \autoref{fig:vip_MeetupNp_CODLR_CCP_gw_H_MFN} and \autoref{fig:vip_MeetupNp_CODLR_CCP_gw_TFN}), 186 common egos) datasts.



    \clearpage
\section{Time lag effect}\label{sec:time_lag_effect}

\begin{figure}[htbp]
	\centering
	\includegraphics[width=0.99\linewidth]{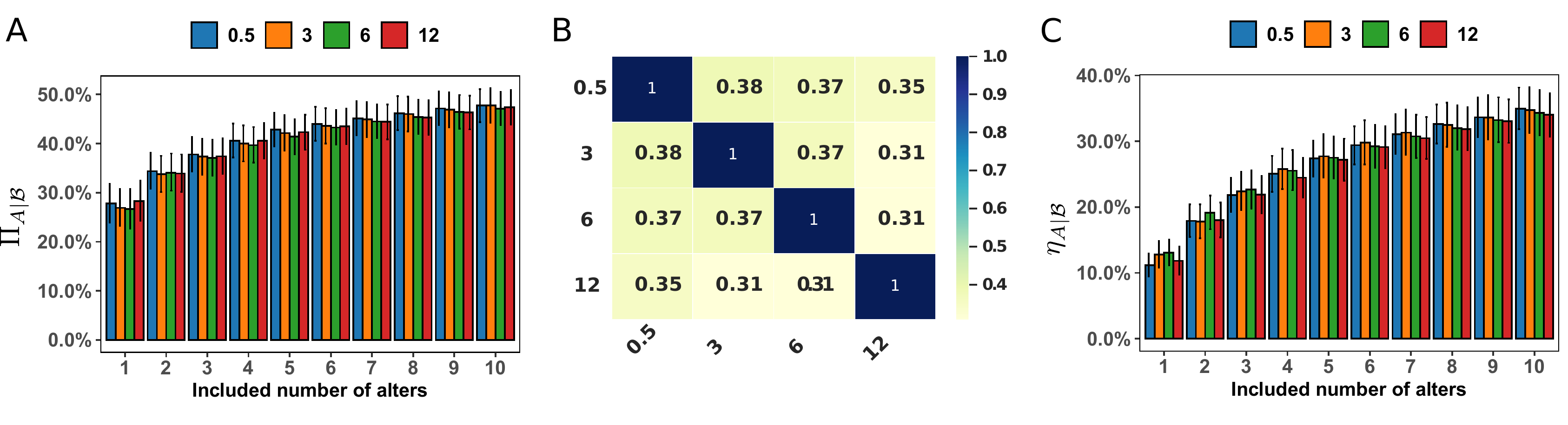}
	\caption{ \textbf{The comparison among the non-social co-located alters within 0.5H, 3H, 6H, 12H one hour sliding windows in Weeplaces dataset}.
	\textbf{}A, (Cumulative) cross-predictability $\Pi_{A|\mathcal{B}}$ VS different included number of alters.
	\textbf{B}, Global Jaccard Similarity between the the non-social co-located alters within 0.5H, 3H, 6H, 12H one hour sliding windows.
	\textbf{C}, (Cumulative) overlapped distinct location ratio $\eta_{A|\mathcal{B}}$ VS different included number of alters.
    Error bars denote mean $\pm95\%$ CI.
	}
	\label{fig:time_effect_combine}
\end{figure}

To check the similarity in pair-wise connections of the different temporal-lag networks, we compute and Jaccard similarity defined for any two sets $A,B$ as
$
J(A, B) = \frac{|A \cap B|}{|A \cup B | },
$
where $| \cdot |$ is the number of elements in the set. All ego-alter pairs in each type of network are considered the sets $A$ and $B$. The results for selected temporal-lag networks are presented in \autoref{fig:time_effect_combine}. The $T$Hr-lag networks correspond to sliding windows where an ego check-in at time $t$ colocates with an alter on the interval $(t-T,t-(T-.5))\bigcup((t+(T-.5),t+T)$. This means a $.5$Hr-lag network corresponds to a 1Hr sliding window colocation network with no temporal lag. The ego-alter pairs of all temporal-lag networks are generally different but provide similar trends in cross predictability and in cumulative overlapped distinct locations.

